\documentclass[11pt]{article}
\usepackage[a4paper,margin=1in]{geometry}
\usepackage[utf8]{inputenc}
\usepackage[T1]{fontenc}
\usepackage{mathptmx}
\usepackage{microtype}
\usepackage{amsmath}
\usepackage{graphicx}
\usepackage{booktabs}
\usepackage{longtable}
\usepackage{array}
\usepackage{enumitem}
\usepackage[section]{placeins}
\usepackage{xcolor}
\usepackage{tikz}
\usetikzlibrary{arrows.meta,positioning,fit,calc,shapes.misc}
\usepackage[font=small,labelfont=bf,labelsep=period]{caption}
\usepackage[round,authoryear]{natbib}
\usepackage[colorlinks=true,linkcolor=blue!50!black,citecolor=blue!50!black,urlcolor=blue!50!black]{hyperref}
\usepackage[nameinlink]{cleveref}
\newcolumntype{P}[1]{>{\raggedright\arraybackslash}p{#1}}
\setlist[enumerate]{itemsep=1pt,topsep=2pt}

\crefname{figure}{Fig.}{Figs.}
\Crefname{figure}{Fig.}{Figs.}
\crefname{section}{Section}{Sections}
\crefname{subsection}{Section}{Sections}
\crefname{subsubsection}{Section}{Sections}
\Crefname{subsection}{Section}{Sections}
\Crefname{subsubsection}{Section}{Sections}
\Crefname{section}{Section}{Sections}
\crefname{table}{Table}{Tables}
\Crefname{table}{Table}{Tables}
\crefname{appendix}{Appendix}{Appendices}
\Crefname{appendix}{Appendix}{Appendices}
\newcommand{\HT}{HydroTuring}
\newcommand{\codeus}{\textunderscore\allowbreak}
\newcommand{\code}[1]{\texttt{\hyphenchar\font=`\-\relax\let\_\codeus #1}}

\title{\HT: A physics probe benchmark testing mass, energy and momentum conservation in hydrologic models}
\author{}
\date{}

\begin{document}

{\centering
{\LARGE\bfseries \HT: A physics probe benchmark testing mass, energy and momentum conservation in hydrologic models\par}
\vspace{1.2em}
}

\begin{abstract}\noindent Hydrologic models, including deep learning systems now deployed for operational flood forecasting, are evaluated primarily by how closely their simulated hydrographs match observed streamflow. However, high hydrograph skill does not guarantee physical consistency: in a synthetic calibration experiment, a conceptual model that loses 15\% of its water input to an unreported sink still reaches a Nash-Sutcliffe efficiency of 0.93, and no shared benchmark tests whether models obey fundamental conservation laws. Here we introduce \HT{} (v0.1.0), a black-box benchmark of 37 physics probes that test mass (24 probes), energy (7) and momentum (6) conservation in hydrologic models. Each probe runs a controlled experiment on forcing generated at run time from a recorded seed and applies binary pass or fail criteria validated against 18 physically exact and 59 deliberately broken reference models. Across 32 evaluated models (3 exact references, 7 AI-based models and 22 process-based models), only the 3 exact references pass every probe among the 28 models scored on 5 or more probes. Process-based and AI-based models pass similar shares of their scored probes (76\% and 74\%), and a pre-trained flood-forecasting LSTM fails 6 of the 11 runoff-only probes it can be scored on. Furthermore, 49\% of all 1{,}184 model-probe pairs remain unscored because models omit the required state or flux variables, while the scored failures occur in perturbation responses and state bounds rather than long-term water-budget closure. \HT{} provides a reproducible smoke test for model development, peer review and operational deployment under an open community protocol.
\end{abstract}

\begin{center}
\small
\textbf{Website} \href{https://flood-lab.github.io/HydroTuring/}{flood-lab.github.io/HydroTuring}
\quad$\vert$\quad
\textbf{GitHub} \href{https://github.com/Flood-Lab/HydroTuring}{github.com/Flood-Lab/HydroTuring}
\quad$\vert$\quad
\textbf{Zenodo} \href{https://doi.org/10.5281/zenodo.23874854}{doi:10.5281/zenodo.23874854}
\end{center}

\begin{center}
{\large\bfseries Authors}\par\smallskip
\par\smallskip\noindent{\small\textsc{Maintainer}}\\ Zhi Li\textsuperscript{1*}
\par\smallskip\noindent{\small\textsc{Probe authors, in order of first merged probe}}\\ Changming Li\textsuperscript{2}, Siavash Shams\textsuperscript{3}, Han Wang\textsuperscript{4}, Qingyi Yang\textsuperscript{5}, Yuanhang Liu\textsuperscript{6}, Taiqi Lian\textsuperscript{7}, Xin Lan\textsuperscript{8}, Shunan Zhou\textsuperscript{9}, Binlan Zhang\textsuperscript{10}, Siddik Barbhuiya\textsuperscript{11}, Songkun Yan\textsuperscript{12}, Bing Li\textsuperscript{6}, Jinlong Hu\textsuperscript{13}, Yaji Wang\textsuperscript{14}, Zhenjiang Wu\textsuperscript{15}, Mofan Zhang\textsuperscript{16}, Kaihao Long\textsuperscript{17}, Jingzhi Chen\textsuperscript{18}
\par\smallskip\noindent{\small\textsc{Model contributor}}\\ Yuan-Heng Wang\textsuperscript{19}
\par\medskip
\begin{minipage}{0.92\linewidth}\raggedright\footnotesize
\textsuperscript{1}Flood Lab, Department of Civil, Environmental and Architectural Engineering, and Institute of Arctic and Alpine Research, University of Colorado Boulder, Boulder, CO, USA\par
\textsuperscript{2}South China University of Technology, China\par
\textsuperscript{3}Department of Electrical Engineering, Columbia University, New York, NY, USA\par
\textsuperscript{4}Department of Civil and Environmental Engineering, The Hong Kong University of Science and Technology, Hong Kong, China\par
\textsuperscript{5}Department of Civil and Environmental Engineering (D.I.C.A.), Politecnico di Milano, Milano 20133, Italy\par
\textsuperscript{6}Independent researcher\par
\textsuperscript{7}Laboratory of Catchment Hydrology and Geomorphology, \'Ecole Polytechnique F\'ed\'erale de Lausanne (EPFL), 1951 Sion, Switzerland\par
\textsuperscript{8}Center for Systems Integration and Sustainability, Department of Fisheries and Wildlife, Michigan State University, East Lansing, MI 48823, USA\par
\textsuperscript{9}Dalian University of Technology, Dalian, China\par
\textsuperscript{10}Institute of Mountain Hazards and Environment, Chinese Academy of Sciences, Chengdu, China\par
\textsuperscript{11}Indian Institute of Technology Mandi, Mandi, Himachal Pradesh, India\par
\textsuperscript{12}School of Civil Engineering and Environmental Science, University of Oklahoma, Norman, OK, USA\par
\textsuperscript{13}State Key Laboratory of Earth Surface Processes and Disaster Risk Reduction, Faculty of Geographical Science, Beijing Normal University, Beijing, China\par
\textsuperscript{14}University of Illinois Urbana-Champaign, Urbana, IL, USA\par
\textsuperscript{15}Institute of Global Environmental Change, Xi'an Jiaotong University, Xi'an 710049, China\par
\textsuperscript{16}Department of Civil and Environmental Engineering, Stanford University, Stanford, CA, USA\par
\textsuperscript{17}School of Geography and Planning, Sun Yat-sen University, Guangzhou 510006, China\par
\textsuperscript{18}Department of Computer Science and Engineering, State University of New York at Buffalo, Buffalo, NY, USA\par
\textsuperscript{19}University of Arizona, Tucson, AZ, USA\par
\par\smallskip\textsuperscript{*}Corresponding author: Zhi Li, \texttt{Zhi.Li-2@colorado.edu}.
\end{minipage}
\end{center}

\vspace{0.6em}

\bigskip
\section{Introduction}\label{sec:intro}

Machine learning has changed hydrologic modeling. Deep networks trained on large-sample data now reproduce runoff with skill that matches or exceeds process-based models \citep{kratzert2018,kratzert2019,nearing2021}. These models are not constrained by the laws of mass, energy and momentum conservation \citep{reichstein2019,karniadakis2021}. Their predictions can become unphysical under out-of-sample conditions such as extremely high or low streamflow \citep{frame2022}. Even in a simple conceptual model, a severe conservation failure can hide behind a high hydrograph score.

In a synthetic calibration experiment, we fitted 2 conceptual models to the same daily runoff record. Model A (\code{reference\_bucket}) conserves water, while model B (\code{reference\_leaky}) sends 15\% of its water input to a sink that its outputs never report. Against the observed streamflow, their Nash-Sutcliffe efficiencies (NSE) are 0.97 and 0.93. Both exceed the 0.75 threshold that a widely cited guideline rates as very good \citep{moriasi2007}. A hydrograph score cannot tell the 2 apart, although model B loses 15\% of every year's precipitation. This is the long-standing problem of getting the right answer for the wrong reason \citep{kirchner2006}, and it is a case of equifinality \citep{beven2006}.

\begin{figure}[!tbp]
\centering
\includegraphics[width=\linewidth]{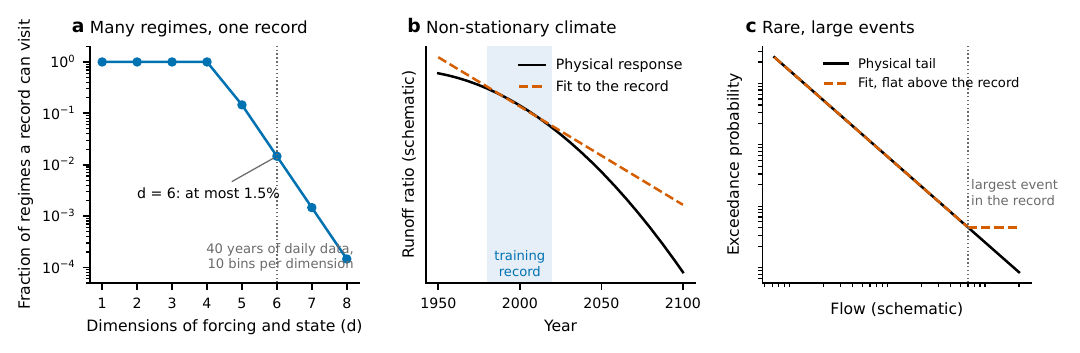}
\caption{Why a streamflow fit does not establish physics. \textbf{a}, Upper bound on the fraction of distinct regimes that a record can visit, for 14{,}610 daily samples and 10 bins per dimension. The count is arithmetic, not data. \textbf{b}, \textbf{c}, Schematics of 2 failure modes of a pure streamflow fit. A fit trained on a record departs from a drifting physical response (\textbf{b}), and it stays flat above the largest event in the record (\textbf{c}). Panels \textbf{b} and \textbf{c} are illustrations and carry no data.}\label{fig:concept}
\end{figure}

The problem grows with the complexity of the model. A hydrologic system is driven by precipitation, temperature and evaporative demand, and it carries soil, snow and groundwater states. With 6 such dimensions and 10 bins per dimension, a 40-year daily record of 14{,}610 samples can visit at most 1.5\% of the distinct regimes (\cref{fig:concept}a). This is the curse of dimensionality \citep{bellman1957}. A model fitted to streamflow alone is therefore constrained on a thin slice of the space it will later be asked to cover. Relying on historical streamflow alone exposes a model to 3 failure modes. First, climate is not stationary \citep{milly2008}, so a response fitted to the record drifts away from the physical response as conditions leave it (\cref{fig:concept}b). Second, the largest events are rare and carry little weight in squared-error scores \citep{gupta2009,clark2021}, so a fitted model can flatten above the largest flow it has seen (\cref{fig:concept}c). Third, a model that gains or loses water at each step accumulates the imbalance in its stores, and the error grows over a long simulation without showing in the streamflow \citep{hoedt2021}. These limits are not unique to AI models, and process-based models can also fail due to imperfect mass closure. However, deep learning rainfall-runoff models tend to degrade more than well-calibrated process-based models on events beyond their training record \citep{frame2022}. \HT{} therefore targets hydrologic models as a whole, whether learned, physical or hybrid, because a physical model can also fail a conservation test.

Current evaluation standards rely on NSE, the Kling-Gupta efficiency (KGE) and related scores on observed basins \citep{nash1970,gupta2009}. Large-sample datasets made such comparisons routine \citep{newman2015,addor2017,kratzert2023}. These scores do not test conservation, the response to a perturbation, or behavior outside the record. Physics-informed learning builds constraints into training \citep{karniadakis2021,hoedt2021,beucler2021}, but no shared test checks whether the trained model honors them. Benchmarks in AI for science report accuracy against public data \citep{rasp2024,takamoto2022}. To our knowledge, no benchmark in hydrology or in Earth system science tests conservation laws on hidden, generated cases. \HT{} is built to fill this gap.

We propose to treat physics as a test suite. The question changes from how close a model comes to 1 record to what the model must do under any record. The name follows Turing's imitation game \citep{turing1950}. That game judges a machine by its behavior under questions it was not prepared for. \HT{} asks whether a model behaves like the physical system under questions that no training record answered. We envision that new AI-based and physics-based models will be tested against \HT{} as a smoke test before they are deployed. 

Here we introduce \HT{} v0.1.0, a probe benchmark for the conservation of mass, energy and momentum in hydrologic models. We define a probe as a controlled experiment with pass or fail criteria on 1 physical law. Our objectives are (1) to define a probe-based benchmark for 3 conservation laws, (2) to run it through a black-box harness that hides the forcing and the tolerances from the model, and (3) to validate every probe against reference models that must pass or must fail. We then report a catalog of 37 probes with the verdicts of 32 models, and we describe an open community protocol that credits probe authors as co-authors. The paper is organized as follows. \Cref{sec:related} reviews related work. \Cref{sec:methods} presents the methods: the design principles (\Cref{sec:design}), the 3 laws and the probe catalog (\Cref{sec:laws}), the harness and its protocol (\Cref{sec:harness}), and the validation of probes against reference models (\Cref{sec:refs}). \Cref{sec:results} reports the verdicts, and \Cref{sec:community} presents the community protocol. \Cref{sec:discussion,sec:conclusion} discuss limitations and conclude. We hope that this release attracts more contributors and users from the community.

\section{Background and related work}\label{sec:related}

\subsection{AI-based hydrologic models}
Data-driven rainfall-runoff modeling has been studied for decades. Artificial neural networks were applied to rainfall-runoff and streamflow forecasting in the 1990s \citep{hsu1995,asce2000,maier2000}. Decision and model trees followed \citep{solomatine2003}, and then random forests and related ensembles \citep{breiman2001,shortridge2016}. These models were fitted to 1 basin at a time and used as time-series predictors. Long short-term memory networks \citep{hochreiter1997} carry memory of past states and predict streamflow with high skill when they are trained on many catchments at once \citep{kratzert2018,kratzert2019}. The same approach now supports forecasts of extreme floods in ungauged watersheds at global scale \citep{nearing2024}. Differentiable models take a different route. They embed a process-based core in a trainable pipeline, so that the physical stores and fluxes remain explicit \citep{feng2022,song2025}. Large language models have also entered the workflow, for example to assist the calibration of process-based models \citep{li2026large}. In practice, all of these approaches are evaluated mainly by hydrograph skill on observed basins. The role of hydrologic science in the age of machine learning is an open question \citep{nearing2021}. A review of data-driven Earth system science argued for hybrid approaches that combine learning with process understanding, and named physical consistency and extrapolation as open challenges \citep{reichstein2019}.

\subsection{Physics in machine learning}
Physics-based (physics-informed or physics-guided) AI has become prevalent in recent years. Physics-informed neural networks (PINN) add the governing equations to the training loss \citep{raissi2019,karniadakis2021}. Graph neural networks can embed the shallow water equations for flood prediction, as in HydroGraphNet \citep{taghizadeh2025}. Hard-constrained networks enforce analytic constraints by construction \citep{beucler2021}. Mass-conserving architectures route water through explicit stores, either inside a recurrent cell \citep{hoedt2021} or as a perceptron with gates \citep{wang2024mcp}. Theory-guided data science frames the broader aim \citep{karpatne2017}. These methods build physics into training. \HT{} tests it after training, on the model as delivered, without access to its weights. The 2 are complementary. Most of these methods target mass, while \HT{} probes energy and momentum conservation as well. A mass-conserving architecture should pass the closure probes by construction, and the energy and momentum probes show what construction alone does not guarantee (\cref{sec:results}).

\subsection{Evaluation practice in hydrology}
The Nash-Sutcliffe efficiency \citep{nash1970} and the Kling-Gupta efficiency \citep{gupta2009} are the standard scores. Both aggregate errors over space and time. Neither checks the physics inside the model, so a model can give the right answer for the wrong reason \citep{kirchner2006}, and models with different internal budgets can earn the same score \citep{beven2006}. A score earned on 1 record also says little about transferability to other basins, periods and climates. Guidelines rate model performance by threshold values of these scores \citep{moriasi2007}, and the common metrics are open to misuse when their assumptions are ignored \citep{clark2021}. Closure of the mass, energy and momentum balances is traditionally a self-check by the model developer. It is rarely reported in a comparable form, and it stays hidden when the model is not open source. The method of multiple working hypotheses aims to expose such structural differences \citep{clark2011}. The central concern is whether a model gives the right answer for the right reasons. \HT{} tests the reasons directly, and it needs no access to the source. It asks what the model does when 1 input changes and everything else is held fixed.

\subsection{Benchmarks in hydrology and AI for science}
Large-sample datasets such as CAMELS \citep{newman2015,addor2017} and Caravan \citep{kratzert2023} made reproducible comparisons across hundreds of basins routine. WeatherBench~2 does the same for global weather forecasts \citep{rasp2024}. PDEBench provides simulated datasets for scientific machine learning and reports, among its metrics, the deviation of predictions from conserved quantities \citep{takamoto2022}. Agents' Last Exam is a living benchmark whose task pool grows as new workflows are onboarded, with verifiable outcomes \citep{sun2026ale}. The community model intercomparison projects, of which CMIP is the best known, show how a shared protocol can be extended by many groups \citep{eyring2016}. \Cref{tab:related} compares these efforts with \HT{} on 4 properties. To our knowledge, none of these efforts tests physical laws with generated cases and hidden tolerances.

\begin{table}[!tbp]
\centering\small
\caption{Comparison of \HT{} with related benchmarks. Entries describe what each effort reports in its primary publication. ``Generated'' means that test cases are produced at run time from a seed, so that no fixed test set exists.}\label{tab:related}
\begin{tabular}{@{}P{0.2\linewidth}P{0.22\linewidth}P{0.2\linewidth}P{0.3\linewidth}@{}}
\toprule
Benchmark & Quantity evaluated & Test cases & Physical laws tested \\ \midrule
CAMELS, Caravan \citep{newman2015,addor2017,kratzert2023} & Streamflow skill in observed basins & Public observed records & Not tested \\
WeatherBench~2 \citep{rasp2024} & Forecast skill against reanalysis and observations & Public data & Not tested as laws \\
PDEBench \citep{takamoto2022} & Accuracy of learned PDE solvers & Public simulated datasets & Partly, through the error of conserved quantities \\
Agents' Last Exam \citep{sun2026ale} & Agent performance on professional workflows & Tasks with verifiable outcomes, extended continuously & Not applicable \\
CheckList \citep{ribeiro2020} & Behavior of NLP models & Test templates & Not applicable \\
\HT & Whether a model conserves mass, energy and momentum & Generated at run time from a seed, with hidden tolerances & Mass, energy and momentum, one law per probe \\ \bottomrule
\end{tabular}
\end{table}

\subsection{Behavioral testing}
Behavioral testing evaluates a model through the outputs it gives to controlled inputs, without access to its internals. CheckList applies this idea to NLP models, with templates that generate invariance and directional expectation tests \citep{ribeiro2020}. \HT{} applies the same idea to physical laws. A probe changes one input, such as air temperature, storm depth or the time step, and states what the output must do. Warmer air over the same rain must lower runoff, and the same water delivered at a finer time step must give the same volume.

\section{Methods}\label{sec:methods}

\subsection{Design principles}\label{sec:design}

The design of \HT{} follows 7 principles. Each sets a standard for how a model should be tested to ensure it respects the underlying physics.

\paragraph{(1) Physical Probes.}
A probe asks 1 physical question and returns a verdict. The question is stated before the test is built, for example whether the same weather delivered at a finer time step gives the same runoff volume over time. A probe returns PASS, FAIL or N/A, and each criterion inside it keeps its own number. The N/A is applied when a particular model cannot be tested for that criterion. For example, if a model does not support sub-daily inputs, the \code{resolution\_invariance} criterion would be marked N/A. There is no weighted aggregate of probes. A score of 0.8 on a composite cannot say which law a model violates, while a failed \code{resolution\_invariance} criterion identifies both the violated law and the error magnitude.

\paragraph{(2) Hidden contract.}
A model is shipped as a Docker container with a manifest and a thin adapter, in any language, and it runs without network access. The harness writes a request, the forcing and the static attributes to a read-only directory, runs the entrypoint once, and reads the outputs from a result file. The model never sees the probe code or the tolerances. This keeps the benchmark open to learned models, process-based models and hybrids on equal terms, and it removes any need to share weights or source.

\paragraph{(3) Hidden forcing.}
A model is tested on cases it cannot have seen. Each probe generates its forcing at run time from a recorded seed. No dataset is committed to the repository, so there is nothing to memorize or tune against. Each probe runs several seeds, and all must pass. Of the 37 probes, 22 use 3 seeds, 14 use 5, and 1 uses 12. Seeds are recorded, so that any failure can be reproduced. A private suite can be added from outside the repository without changing the public one, which allows an operator to keep further probes undisclosed.

\paragraph{(4) Paired cases.}
Many physical laws constrain how a system responds to a perturbation rather than a single state value. Warmer air over the same rain must lower runoff, and a storm added on a given day cannot change anything before that day. A probe therefore runs a control and 1 perturbed copy of the same case, which differ in 1 input, and it compares the 2 runs. This checks response and invariance, which an absolute closure test cannot reach. A model that ignores an input passes every single-run criterion, and a paired probe is the only test that catches it. Of the 37 probes, 18 compare 2 or more runs of the same case.

\paragraph{(5) Labeled regimes.}
Generators label the cases they produce. A label column, such as the stretch of a snow season or the days of an isolated storm, is read by the harness and never passed to the model. Criteria use the labels to score a regime separately, so that errors of opposite sign in different regimes cannot cancel, and the report can trace a failure to the regime that produced it.

\paragraph{(6) Strict aggregate.}
A model passes only when every criterion of every probe that applies to it passes. A single violated law is enough to fail. We adopt this strict rule because physical conservation is binary under a defined tolerance. A model that leaks water in 1 regime fails to conserve water, even if it closes the budget in all others.

\paragraph{(7) Validated probes.}
A probe is only as good as its ability to separate right from wrong. Every probe must therefore pass an acceptance gate before it is merged. The gate runs exact reference models, which must pass, and deliberately broken reference models, which must fail and must trip the criterion that each was built to trip. If a probe fails an exact model, its tolerance has drifted beyond what exact physics achieves, and the probe is wrong. If a probe passes a broken model, it does not discriminate. We return to the gate in \cref{sec:refs}.

\subsection{The three laws and the probe catalog}\label{sec:laws}

\begin{figure}[!tbp]
\centering
\includegraphics[width=\linewidth]{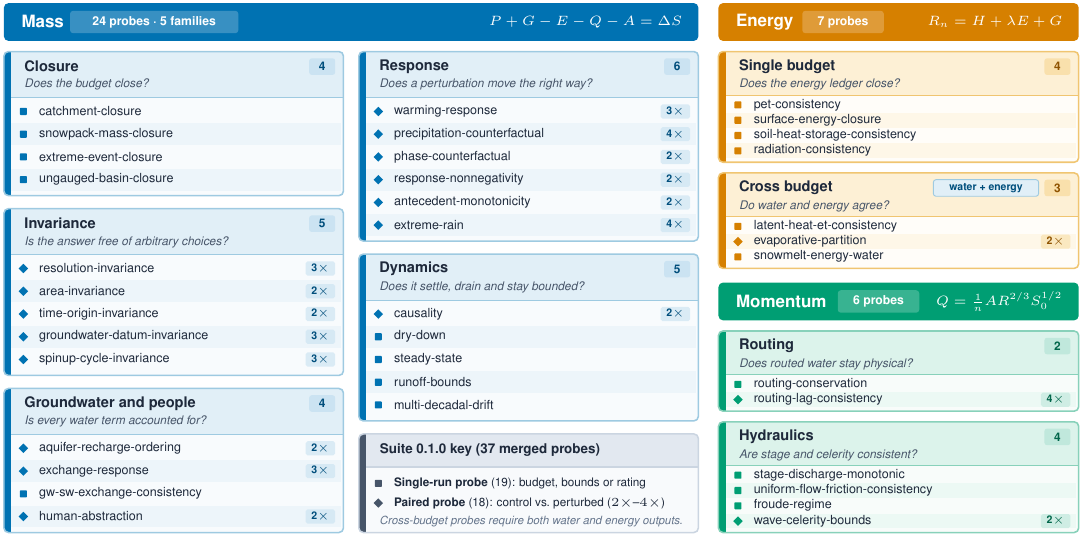}
\caption{Map of the 37 merged probes by law and family. Each entry is 1 probe in suite version 0.1.0. The cross-budget family needs outputs from both the water and the energy ledgers.}\label{fig:map}
\end{figure}

\HT{} version 0.1.0 holds 37 probes, merged between 3 September and 16 October 2026: 24 for mass, 7 for energy and 6 for momentum. The probes use 44 distinct criterion types. We group them into 9 families (\cref{fig:map,tab:families}). \Cref{app:catalogue} lists every probe with its question, its criteria and its authors.

\subsubsection{Mass}
The governing relation is the water budget of the control volume,
\begin{equation}
P + G - E - Q - A = \Delta S_{\mathrm{soil}} + \Delta S_{\mathrm{snow}} + \Delta S_{\mathrm{canopy}} + \Delta S_{\mathrm{gw}} + \Delta S_{\mathrm{ch}},
\end{equation}
where $P$ is precipitation, $E$ is evapotranspiration including sublimation, $Q$ is runoff, $G$ is an exchange that the model declares with the outside, $A$ is a prescribed human water withdrawal, and $\Delta S$ is the change in each reported store. The 24 mass probes fall into 5 families.

\textbf{Closure probes} ask whether the budget closes. The reference probe, \code{catchment-closure}, forces a lumped catchment with 10 years of daily weather and 5 seeds. It requires the cumulative residual to stay within 5\% of cumulative precipitation, every store to remain within its physical range, and evaporation to stay at or below demand. The other 3 closure probes close the budget inside the snowpack across accumulation, storage and melt (\code{snowpack-mass-closure}), event by event under rainfall that is scaled toward 100-year depths (\code{extreme-event-closure}), and across 12 independent catchments with sampled soil and canopy capacities (\code{ungauged-basin-closure}).

\textbf{Invariance probes} ask whether the answer depends on how the modeler set up the domain. A model of the same physical system, discretized at a different spatial or temporal resolution, should give the same result. Each probe changes 1 such setup choice and checks that the output does not move:
\begin{enumerate}[itemsep=1pt,topsep=2pt,leftmargin=1.8em]
\item The same month of weather delivered at 1-minute, hourly and daily steps must give the same runoff volume within 10\% (\code{resolution-invariance}).
\item The same weather on a catchment 10 times larger must give the same depths (\code{area-invariance}).
\item A 28-year shift of the calendar must leave fluxes and stores unchanged to a relative $10^{-9}$ (\code{time-origin-invariance}).
\item A common 250~m shift of all groundwater heads must leave the exchange unchanged (\code{groundwater-datum-invariance}).
\item A periodic climate must forget extra spin-up cycles (\code{spinup-cycle-invariance}).
\end{enumerate}

\textbf{Response probes} ask whether a perturbation moves the output in the right direction and by a physical amount. Each probe applies 1 change to the forcing or the initial state and checks the response against a physical bound:
\begin{enumerate}[itemsep=1pt,topsep=2pt,leftmargin=1.8em]
\item Under identical precipitation, air 3~$^\circ$C warmer must lower runoff and air 3~$^\circ$C cooler must raise it, each by at least 0.1 of the change in evaporative demand (\code{warming-response}).
\item Rain added to a run must appear in the budget and be shared among evaporation, runoff and storage (\code{precipitation-counterfactual}).
\item Rain that would have fallen as snow must yield the same water within 5\% (\code{phase-counterfactual}).
\item A 120~mm storm added on 1 day may not lower runoff on any later day (\code{response-nonnegativity}).
\item The same 60~mm storm on a wet catchment must run off more than on a dry one, by at least 2\% of the storm and by no more than the 120~mm of extra water (\code{antecedent-monotonicity}).
\item Scaling the largest storm to 10 times its size must not lower runoff, must not add more runoff than rain, and must send at least 10\% of the added rain to runoff (\code{extreme-rain}).
\end{enumerate}

\textbf{Dynamics probes} ask whether the catchment settles, drains and stays bounded. Each probe tests 1 part of the catchment's behavior, from the moment of the storm to 50 years of repeated weather:
\begin{enumerate}[itemsep=1pt,topsep=2pt,leftmargin=1.8em]
\item A storm may not change anything before it falls (\code{causality}).
\item After the rain stops, runoff and every store may only decline, and the total that runs off cannot exceed what the catchment held (\code{dry-down}).
\item Constant weather for 3 years must produce constant stores and runoff within 1\% (\code{steady-state}).
\item Over 10 years, runoff must lie between rain minus potential evaporation minus capacity and rain plus the water the stores could release (\code{runoff-bounds}).
\item Repeated weather over 50 years must not show storage drift beyond $2\times10^{-4}$ of precipitation over the final 5-year block (\code{multi-decadal-drift}).
\end{enumerate}

\textbf{Groundwater and human probes} ask whether every water term is accounted for. Each probe checks 1 water term that the model declares or prescribes against the budget:
\begin{enumerate}[itemsep=1pt,topsep=2pt,leftmargin=1.8em]
\item Direct aquifer recharge must preserve storage and the ordering of exchange (\code{aquifer-recharge-ordering}).
\item A declared head-driven exchange must respond to its external head (\code{exchange-response}).
\item In a 2-year aquifer case, a model that reports groundwater-river exchange must account for recharge, both exchange directions and aquifer storage, so that exchange and storage close the water budget (\code{gw-sw-exchange-consistency}).
\item A prescribed irrigation withdrawal of 380~mm over 10 years must leave the budget, and the difference between the natural and the irrigated run must equal the abstracted volume within 5\% (\code{human-abstraction}).
\end{enumerate}

\subsubsection{Energy}
The governing relation is the surface energy budget,
\begin{equation}
R_n = H + \lambda E + G, \qquad \lambda E = \lambda(T)\,E,
\end{equation}
where $R_n$ is net radiation, $H$ is the sensible heat flux, $\lambda E$ is the latent heat flux, $G$ is the ground heat flux and $\lambda(T)$ is the temperature-dependent latent heat of the phase change. Seven probes test the energy budget.

Four single-budget probes test whether the energy balance holds by itself: (1) Evaporation on wet days must follow demand and on dry days must follow water (\code{pet-consistency}). (2) Hourly residuals must close within each solar day or night, with absolute residuals averaged so that opposite errors cannot cancel (\code{surface-energy-closure}). (3) Heat entering a soil layer minus heat leaving it must equal the change in stored heat (\code{soil-heat-storage-consistency}). (4) Upward longwave radiation must equal emission at the skin temperature plus reflected downward longwave, within 0.5\% or 0.5~W~m$^{-2}$ (\code{radiation-consistency}).

Three cross-budget probes test whether the water and energy budgets agree: (1) A model with a water head and an energy head can close both budgets and still contradict itself, because the inconsistency lies in the coupling between the two budgets. Latent heat must equal evapotranspiration times the latent heat of the phase it underwent, within 0.5\% (\code{latent-heat-et-consistency}). (2) A drying surface under fixed net radiation must give up its latent heat to the sensible and ground fluxes, and the 3 changes must sum to zero (\code{evaporative-partition}). (3) Melt that the water budget reports must be paid for in the energy budget, by the latent heat of fusion and the cold content of the pack (\code{snowmelt-energy-water}).

\subsubsection{Momentum}
Momentum conservation in a channel is expressed by the Saint-Venant equations. The suite does not yet test the full unsteady balance. It tests conditions that any solution of the equations must satisfy. In the steady uniform limit the bed slope $S_0$ equals the friction slope $S_f$, and Manning's relation links discharge $Q$ and stage through the channel geometry,
\begin{equation}
S_f = S_0, \qquad Q = \frac{1}{n}\,A\,R^{2/3}\,S_0^{1/2},
\end{equation}
where $n$ is Manning's roughness, $A$ the flow area and $R$ the hydraulic radius. On a mild slope the flow stays subcritical, so the Froude number remains below 1.

Two routing probes test whether routed water stays physical: (1) A routing store may never be negative, and it may never hold more than the cumulative runoff inflow over a 15-day hydrograph (\code{routing-conservation}). (2) The lag of a runoff peak must follow a Snyder travel-time scale \citep{snyder1938} within a factor of 2 across 4 catchments and must grow with catchment size (\code{routing-lag-consistency}). Four hydraulics probes test the consistency of reported stage and discharge: (1) Stage must be single-valued in discharge and rise with it, with a loop of the right orientation for a flood wave (\code{stage-discharge-monotonic}). (2) Discharge and stage on 3 steady plateaus must satisfy the friction balance within 5\% (\code{uniform-flow-friction-consistency}). (3) The flow must remain subcritical on a mild slope (\code{froude-regime}). (4) The celerity of a small pulse, measured between a short and a long reach, must respond to the hydraulic state (\code{wave-celerity-bounds}).

\subsubsection{Coverage and gaps}
Momentum is the least developed law in version 0.1.0. Its 6 probes test consistency conditions and bounds, and no probe yet tests the unsteady momentum balance reach by reach. Three further probes are on the roadmap. Two are in progress: a probe on reach-by-reach mass closure in a branching network, \code{mass/\allowbreak routing-network-closure}, and a probe on the release of meltwater from a cold snowpack, \code{energy/\allowbreak snowpack-ripening}. The third, \code{momentum/\allowbreak channel-routing-mass}, is unclaimed and requires inflow minus outflow minus storage change per reach. An earlier attempt was set aside because the contract then had no variable for water arriving at a reach, so that both sides of the residual came from the model's own outputs. The contract now carries a prescribed inflow, \code{q\_in}, which the wave-celerity probe uses (\cref{tab:lexicon}). We plan to complete version~1 once the suite reaches 50 probes (\cref{sec:community}).

\begin{table}[!tbp]
\centering\small
\caption{Probe families in suite v0.1.0. The last column gives 1 representative experiment and its limit. Full definitions are in \cref{app:catalogue}.}\label{tab:families}
\begin{tabular}{@{}lP{0.17\linewidth}cP{0.27\linewidth}P{0.27\linewidth}@{}}
\toprule
Law & Family & Probes & Representative experiment & Representative criterion and limit \\ \midrule
Mass & Closure & 4 & 10 years of generated weather, 5 seeds & Cumulative residual $\le$ 5\% of precipitation \\
Mass & Invariance & 5 & Same weather at 1~min, 1~h and 1~d & Runoff volumes agree within 10\% \\
Mass & Response & 6 & Largest storm scaled by 2, 5 and 10 & Added runoff share between 0.1 and 1.05 \\
Mass & Dynamics & 5 & 2 years without rain & Runoff and stores only decline \\
Mass & Groundwater and human & 4 & 380~mm irrigation withdrawal & Budget difference accounts for the volume within 5\% \\
Energy & Single budget & 4 & Hourly bare-surface budget & Residual within 5\% or 2~W~m$^{-2}$ \\
Energy & Cross budget & 3 & 4 months of drought at fixed net radiation & Changes in $\lambda E$, $H$ and $G$ sum to zero \\
Momentum & Routing & 2 & 4 large storms followed by dry spells & Store between zero and a 15-day hydrograph \\
Momentum & Hydraulics & 4 & 3 steady plateaus & Friction balance within 5\% \\ \bottomrule
\end{tabular}
\end{table}

\subsection{Harness and protocol}\label{sec:harness}

\subsubsection{Architecture}
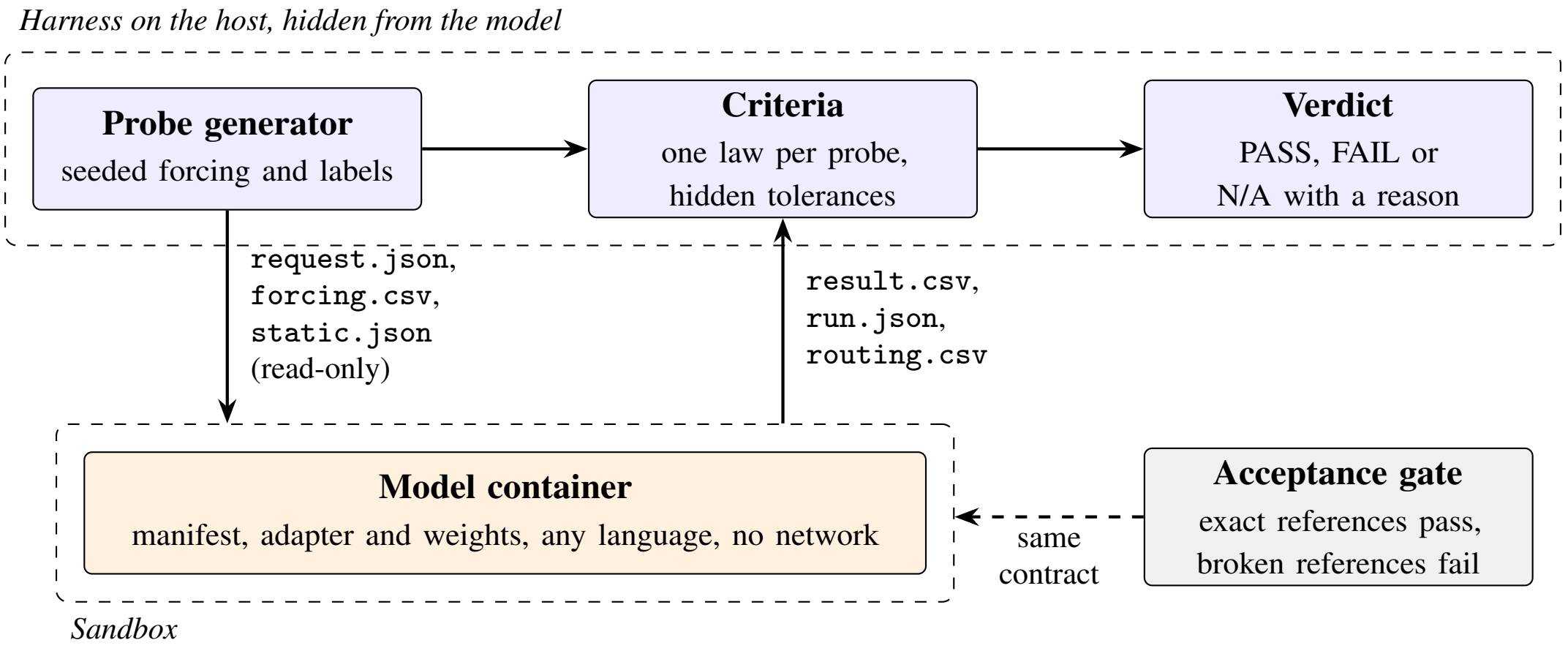
\begin{figure}[!tbp]
\centering
\begin{tikzpicture}[
  font=\footnotesize,
  box/.style={draw, rounded corners=2pt, align=center, minimum height=1.1cm, text width=3.3cm, inner sep=3pt},
  hbox/.style={box, fill=blue!7},
  mbox/.style={box, fill=orange!12, text width=7.4cm},
  gbox/.style={box, fill=gray!12, text width=3.3cm},
  arr/.style={-{Stealth[length=2.2mm]}, thick},
  darr/.style={-{Stealth[length=2.2mm]}, thick, dashed},
  lab/.style={font=\scriptsize, align=left}
]
\node[hbox] (gen) {\textbf{Probe generator}\\[1pt]\scriptsize seeded forcing and labels};
\node[hbox, right=1.5cm of gen] (crit) {\textbf{Criteria}\\[1pt]\scriptsize one law per probe, hidden tolerances};
\node[hbox, right=1.5cm of crit] (ver) {\textbf{Verdict}\\[1pt]\scriptsize PASS, FAIL or N/A with a reason};
\node[mbox] (mod) at ($(gen.south)!0.5!(crit.south)+(0,-2.7)$) {\textbf{Model container}\\[1pt]\scriptsize manifest, adapter and weights, any language, no network};
\node[gbox] (gate) at ($(ver.south)+(0,-2.7)$) {\textbf{Acceptance gate}\\[1pt]\scriptsize exact references pass, broken references fail};
\node[draw, dashed, rounded corners=3pt, fit=(gen)(crit)(ver), inner sep=7pt] (hframe) {};
\node[anchor=south west, font=\scriptsize\itshape] at (hframe.north west) {Harness on the host, hidden from the model};
\node[draw, dashed, rounded corners=3pt, fit=(mod), inner sep=7pt] (mframe) {};
\node[anchor=north west, font=\scriptsize\itshape] at (mframe.south west) {Sandbox};
\draw[arr] (gen) -- (crit);
\draw[arr] (crit) -- (ver);
\draw[arr] (gen.south) -- node[right, lab, xshift=2pt] {\code{request.json},\\\code{forcing.csv},\\\code{static.json}\\(read-only)} (gen.south |- mframe.north);
\draw[arr] (crit.south |- mframe.north) -- node[right, lab, xshift=2pt] {\code{result.csv},\\\code{run.json},\\\code{routing.csv}} (crit.south);
\draw[darr] (gate.west) -- node[below, lab, align=center] {same\\contract} (mframe.east |- gate.west);
\end{tikzpicture}
\caption{Architecture of the \HT{} harness. The probe generator, the criteria and the tolerances stay on the host. The model runs in a sealed container, reads the forcing and static attributes from a read-only directory, and writes the outputs that the criteria score. Reference models enter through the same contract in the acceptance gate (\cref{sec:refs}).}\label{fig:harness}
\end{figure}

The harness separates what a model can see from what decides its verdict (\cref{fig:harness}). It does so by splitting each evaluation between two sides. On the host, a probe generates the forcing and the labels, and it keeps the labels, the criteria and their tolerances to itself. The model runs on the other side, in a sealed container, and receives only the forcing and the static attributes. Because the model never sees the labels or the criteria, it cannot tune its output to the verdict. Two files link the model to the harness. A manifest, \code{model.yaml}, declares what the model needs and what it emits, and a thin adapter, \code{ht\_adapter.py}, translates between this contract and the native model, so that a model needs no change to its own code. Three runners then start the model: Docker, a plain subprocess for models that run natively, and Apptainer for clusters.

\subsubsection{Lexicon and input-output contract}
Model variables are named differently in different models. Some models name discharge (cms) as runoff (\code{q}), while others use different conventions. Additionally, a hydrologic model may output runoff alone, and a land surface model may output every flux and store. The contract therefore rests on a shared lexicon. In suite 0.1.0 it names 29 output variables, 14 forcing variables and 34 static attributes, each with units, a sign convention and a statement of what a value means over a time step (\cref{tab:lexicon,app:lexicon}). Names follow the CMIP conventions where an equivalent exists \citep{eyring2016}. A model declares the outputs it emits and the inputs it needs or uses, and it must not pad that list. A probe declares the outputs, forcing and static attributes it requires. The harness scores a pair only when the two declarations are compatible.

\begin{table}[!tbp]
\centering\small
\caption{Core vocabulary of the \texttt{/io} contract. A model reports a subset of the output variables and declares which. A probe scores a model only if its outputs cover what the probe requires. The full lexicon is in Appendix~\ref{app:lexicon}.}\label{tab:lexicon}
\begin{tabular}{@{}llP{0.58\linewidth}@{}}
\toprule Name & Units & Physical meaning \\ \midrule
\texttt{pr} & mm d$^{-1}$ & Precipitation, echoed exactly as the forcing gave it \\
\texttt{evspsbl} & mm d$^{-1}$ & Evapotranspiration, including sublimation; net of dew and frost where the model nets them \\
\texttt{mrro} & mm d$^{-1}$ & Total runoff \\
\texttt{dis} & m$^3$ s$^{-1}$ & River discharge \\
\texttt{gwex} & mm d$^{-1}$ & Declared exchange with the outside: regional groundwater, inter-basin transfer \\
\texttt{snm} & mm d$^{-1}$ & Liquid delivered from the snow module to the ground, rain passed through included, and not an outflow beside \texttt{mrro} \\
\texttt{hfls} & W m$^{-2}$ & Latent heat flux \\
\texttt{hfss} & W m$^{-2}$ & Sensible heat flux \\
\texttt{hfg} & W m$^{-2}$ & Ground heat flux at the actual soil surface, where a deeper flux is corrected for the heat stored above it \\
\texttt{mrso} & mm & Soil water storage (liquid and ice), reported as an absolute state rather than a tendency \\
\texttt{snw} & mm & Snow water equivalent, identically zero and named in \texttt{zero\_stores} for a model with no snow module \\
\texttt{canopy} & mm & Canopy interception storage, identically zero and named in \texttt{zero\_stores} for a model with no such store \\
\texttt{gw} & mm & Groundwater storage below the soil column \\
\texttt{channel} & mm & Water generated as runoff but not yet released by the model's routing \\
\texttt{stage} & m & Water-surface elevation on the case's datum, read beside the row's mean discharge (depth is \texttt{stage} minus \texttt{bed\_elevation\_m}) \\
\midrule \multicolumn{3}{@{}l}{\textit{Forcing a model receives}} \\
\texttt{pr} & mm d$^{-1}$ & Precipitation \\
\texttt{tas} & $^\circ$C & Near-surface air temperature in Celsius, unlike the kelvin of \texttt{ts} and \texttt{tsoil\_layer} \\
\texttt{pet} & mm d$^{-1}$ & Potential evapotranspiration \\
\texttt{rn} & W m$^{-2}$ & Prescribed net radiation, which the model accepts as the energy input to the reported budget when declared \\
\texttt{abstr} & mm d$^{-1}$ & Prescribed human withdrawal, net of return flow \\
\texttt{q\_\allowbreak{}in} & m$^3$ s$^{-1}$ & Prescribed river or upstream inflow, already a discharge and not multiplied by area or row length \\
\bottomrule
\end{tabular}
\end{table}

A run has 3 steps: the harness stages the inputs, the model runs, and the harness checks what the model returns against the contract. In the first step, the harness writes a request, the forcing and the static attributes to a read-only directory, so that the model cannot alter them. On a river network, the run also reads reach-indexed variables from the routing table. In the second step, the entrypoint runs once and writes 1 row per forcing step to the recorded results, together with a provenance file. Spin-up rows are included in these results but are excluded from scoring. In the third step, the harness compares the results with the request. The adapter must emit the exact number of steps that the request names, so a step count mismatch or a missing declared column is a contract error.

\subsubsection{Verdicts}
A probe returns 1 of 3 verdicts and a reason (\cref{tab:verdicts}). PASS and FAIL are scientific statements. N/A means that the probe does not apply for this particular model. For example, a streamflow-only model is N/A on every budget probe. Rather than violating conservation, the model cannot be evaluated on that law because it omits the required outputs. We keep this case apart from FAIL because the 2 statements say different things about a model. The reason is stored beside the verdict, together with the measured quantity, so that a reader can see by how much a limit was missed.

\begin{table}[!tbp]
\centering\small
\caption{Verdicts and reasons. Only PASS and FAIL enter a model's score.}\label{tab:verdicts}
\begin{tabular}{@{}llcP{0.52\linewidth}@{}}
\toprule
Verdict & Reason & Scored & Meaning \\ \midrule
PASS & OK & yes & Every criterion on every seed is within its limit \\
FAIL & VIOLATION & yes & At least 1 criterion on 1 seed is outside its limit \\
N/A & INCOMPLETE & no & The model does not report the variables that the probe needs \\
N/A & INCOMPATIBLE & no & The model and the probe disagree on the time step, on an input, or on paired-perturbation support \\
-- & ERROR & no & The adapter or the machinery failed, which is operational and makes the run exit with an error code \\ \bottomrule
\end{tabular}
\end{table}

A model passes the suite when at least 1 probe could be put to it and every probe it was scored on passes. This is the strict aggregate of \cref{sec:design}. Because N/A probes count neither for nor against a model, a pass fraction is meaningless without the number of probes scored. We therefore always report both.

\subsubsection{Criteria on the reference probe}
Each probe contains criteria, and each criterion returns a binary result and a measured value. The reference probe, \code{catchment-closure}, runs 10 generated years of daily weather after 365 days of spin-up, on 5 seeds. \Cref{tab:crit} lists its criteria. The closure criterion divides the cumulative residual of the budget by cumulative precipitation, and it uses a 5\% limit as an engineering standard. Section~\ref{sec:refs} justifies this limit with the reference models and a sweep over the tolerance.

\begin{table}[!tbp]
\centering\small
\caption{Criteria of the reference probe \code{mass/\allowbreak catchment-closure}. The last 3 criteria are shared with most mass probes.}\label{tab:crit}
\begin{tabular}{@{}P{0.22\linewidth}P{0.30\linewidth}P{0.30\linewidth}@{}}
\toprule
Criterion & Question & Limit \\ \midrule
\code{closure} & Does precipitation minus evaporation, runoff and the change of every store stay near zero? & Cumulative residual at most 5\% of cumulative precipitation \\
\code{et\_plausible} & Does evaporation stay at or below the demand? & No step exceeds the potential evaporation supplied \\
\code{state\_bounds} & Does every reported store stay in its physical range? & No store outside its range \\
\code{non\_degenerate} & Does the model produce runoff that responds to rain? & Runoff ratio between 0.02 and 0.98 and coefficient of variation at least 0.1 \\
\code{forcing\_fidelity} & Does the model echo the forcing that it was given? & Relative tolerance of $10^{-6}$ \\ \bottomrule
\end{tabular}
\end{table}

\subsection{Reference models and probe validation}\label{sec:refs}

\subsubsection{Reference models}
The suite carries 77 reference models, 18 exact and 59 broken. An exact reference conserves the law that a probe tests, by construction or through a published scheme. Four exact references serve as common baselines. The model \code{reference\_bucket} (model A in \cref{sec:intro}) is a single bucket that conserves water exactly. The models \code{flex\_lumped} and \code{flex\_topo} are 2 conceptual models of the FLEX family \citep{hydrologicmodels}, and \code{sacsma\_snow17} couples the Sacramento soil moisture accounting model with the Snow-17 snow model \citep{sacsmasnow17}. The 14 remaining exact references are written for the probes whose physics they exercise. They include an energy-balance snowpack, a soil column with prescribed heat capacity, a Saint-Venant channel and a lagged-head aquifer cell (\cref{tab:exact}).

Each of the 59 broken references carries 1 defect that the suite must detect (\cref{tab:broken}). We use 6 of them throughout the paper. The model \code{reference\_leaky} (model B in \cref{sec:intro}) hides a silent 15\% sink, so its budget does not close. The model \code{reference\_cheater} solves for storage as whatever balances the budget, which closes the budget by construction and drives storage outside its physical range. The model \code{reference\_degenerate} evaporates all precipitation and produces no runoff. The model \code{reference\_saturating} caps daily runoff at 25~mm, so it is exact for ordinary storms and flat for extreme ones. The model \code{reference\_climatology} emits the seasonal mean of runoff and evaporation whatever the weather. The model \code{reference\_in\_sample} leaks water only outside the range for which it was built. A last reference, \code{reference\_streamflow\_only}, is not broken but limited. It reports discharge only, from a store that does not read the temperature. It is N/A on every budget probe, and the warming-response probe catches it.

\subsubsection{Acceptance gate}
A probe enters the suite only when its gate is satisfied. The gate has 2 kinds of check. A must-pass check requires an exact reference to pass the probe. A must-fail check requires a broken reference to fail the probe and to trip the criterion that it was built to trip. The 37 probes carry 116 must-pass and 78 must-fail checks. We re-ran all 37 gates on 2 November 2026 at the commit that tags suite 0.1.0, and every check behaved as declared.

The gate fixes tolerances from 2 sides. An exact model sets the lower bound, because a tolerance tighter than what exact physics achieves would fail honest models. Over 20 random seeds, the exact bucket, \code{flex\_lumped} and \code{sacsma\_snow17} pass \code{catchment-closure} on all 20, with residuals below $2.3\times10^{-15}$ of precipitation. A broken model sets the upper bound. Over the same seeds, \code{reference\_leaky} fails the closure criterion on 20 of 20. The model \code{reference\_cheater} fails the state-bounds criterion on 20 of 20, and \code{reference\_degenerate} fails the non-degeneracy criterion on 20 of 20 (\cref{tab:seeds}).

The closure tolerance sits between the 2 bounds. It is 5\% of cumulative precipitation, the engineering rule that the probe adopts. Among the 22 evaluated models that report a whole-record residual on \code{catchment-closure}, 21 are below 0.2\% and the largest is 2.5\% (\code{xaj}). Any tolerance between 2.6\% and 15\% therefore separates every evaluated model from \code{reference\_leaky}, whose sink is 15\% by construction (\cref{fig:closure}a). We keep 5\% because it is conventional and because it sits in this empty interval. Any change to it requires the process in \cref{sec:community}, and \cref{app:tolerance} gives the full sweep.

A single case can miss a defect. The model \code{reference\_saturating} fails the extreme-rain criterion on 6 of 10 seeds and passes it on 4. The signal therefore varies from seed to seed. Across the ladder up to 10 times the largest storm, the exact bucket returns between 0.56 and 1.00 of the added rain on every 1 of the 10 seeds. The saturating model returns 0.01 to 0.22 and exceeds the limit of 0.10 on 4 of them. A probe that used 1 seed would pass the saturating model 40\% of the time. The probe uses 3 seeds, and a model passes only when all 3 pass. Of the 120 subsets of 3 among the 10 seeds, 116 contain at least 1 seed that exposes the defect. Across the suite, 22 probes use 3 seeds, 14 use 5, and the probe on catchment attributes uses 12.

\section{Results}\label{sec:results}

We evaluated 32 models on the 37 probes of suite 0.1.0. Three are the physically exact references \code{flex\_lumped}, \code{flex\_topo} and \code{sacsma\_snow17}. Seven are AI-based: an LSTM, the differentiable HBV model \code{dhbv2} \citep{song2025}, and 5 training checkpoints of a mass-conserving perceptron, \code{mcp1} to \code{mcp5} \citep{wang2024mcp}. The remaining 22 are process-based models, from conceptual bucket models to land surface models, a groundwater model and routing models. They include \code{gr4j} \citep{perrin2003}, \code{xaj} \citep{zhao1992}, \code{topmodel} \citep{beven1979}, \code{topkapi} \citep{ciarapica2002}, \code{vic} \citep{liang1994,lohmann1998}, \code{mhm} \citep{samaniego2010}, \code{summa} \citep{clark2015}, \code{raven} \citep{craig2020}, \code{hype} \citep{lindstrom2010}, \code{prms} \citep{markstrom2015}, \code{noahmp} \citep{niu2011}, \code{cwatm} \citep{burek2020}, \code{lisflood} \citep{vanderknijff2010}, \code{modflow6} \citep{langevin2017} and \code{mizuroute} \citep{mizukami2016}. Of the 32 models, 29 entered through model proposals (\cref{sec:community}). Each model was run in its own container with the manifest and adapter in the repository, and each version is recorded in the archive.

\subsection{Verdict matrix}
\begin{figure}[!tbp]
\centering
\includegraphics[width=\linewidth]{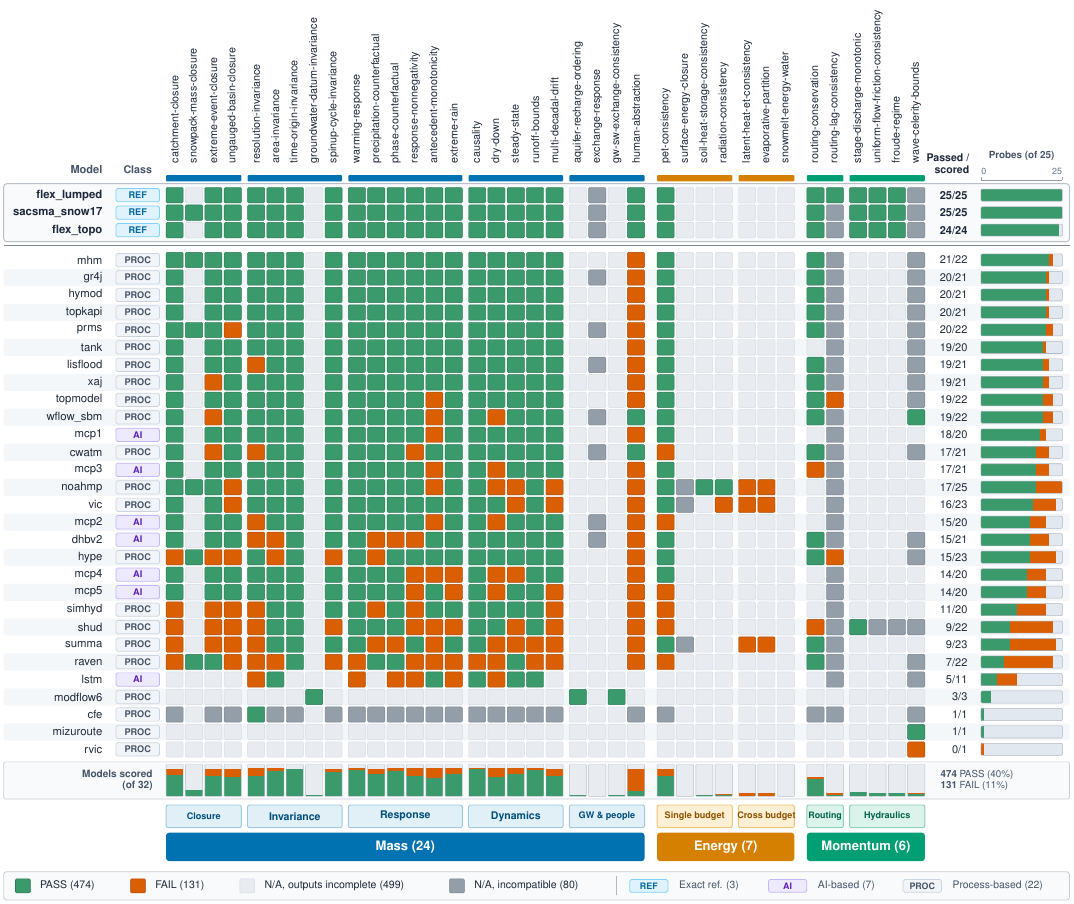}
\caption{Verdict matrix of 32 models on 37 probes in suite 0.1.0. Rows are models. The 3 exact references are on top, and the other models follow in order of the number of probes that they pass. Columns are probes, grouped by law. Green is PASS and orange is FAIL. Light gray is N/A because the model does not report the required variables (INCOMPLETE), and dark gray is N/A because model and probe disagree on an input or a time step (INCOMPATIBLE).}\label{fig:matrix}
\end{figure}

The 32 models and 37 probes give 1{,}184 model-probe pairs (\cref{fig:matrix}). Of these, 474 are PASS, 131 are FAIL and 579 are N/A. Among the N/A pairs, 499 are INCOMPLETE and 80 are INCOMPATIBLE. Consequently, 49\% of all model-probe pairs remain unscored due to missing output variables or time-step mismatches rather than physical failures. Mass probes are N/A for 30\% of pairs (234 of 768), while energy probes are N/A for 84\% (188 of 224) and momentum probes for 82\% (157 of 192). Few models report heat fluxes or reach-indexed variables.

Only the 3 exact references pass every probe that they are scored on (74 of 74 pairs). Among the 28 models that are scored on 5 or more probes, no other model does so. The remaining 4 models, \code{cfe}, \code{mizuroute}, \code{modflow6} and \code{rvic}, are scored on 1 to 3 probes. Three of them pass all of their scored probes, which says little because the sample is small. The 7 AI-based models pass 98 of 133 scored pairs (74\%), while the 22 process-based models pass 302 of 398 (76\%). Because these pass rates are nearly identical, model class alone does not predict physical consistency. \Cref{tab:standings} gives the standing of every model.

Models fail 25 of the 37 probes. Twelve probes fail no evaluated model, and 3 of them, \code{energy/\allowbreak snowmelt-energy-water}, \code{energy/\allowbreak surface-energy-closure} and \code{mass/\allowbreak exchange-response}, have not yet scored any submitted model. The most discriminating probes are \code{mass/\allowbreak human-abstraction} (22 of 27 scored models fail), \code{mass/\allowbreak antecedent-monotonicity} (10 of 28), \code{mass/\allowbreak dry-down} (9 of 28), \code{mass/\allowbreak resolution-invariance} (9 of 29) and \code{mass/\allowbreak response-nonnegativity} (8 of 28).

\subsection{Where models fail}
\begin{figure}[!tbp]
\centering
\includegraphics[width=\linewidth]{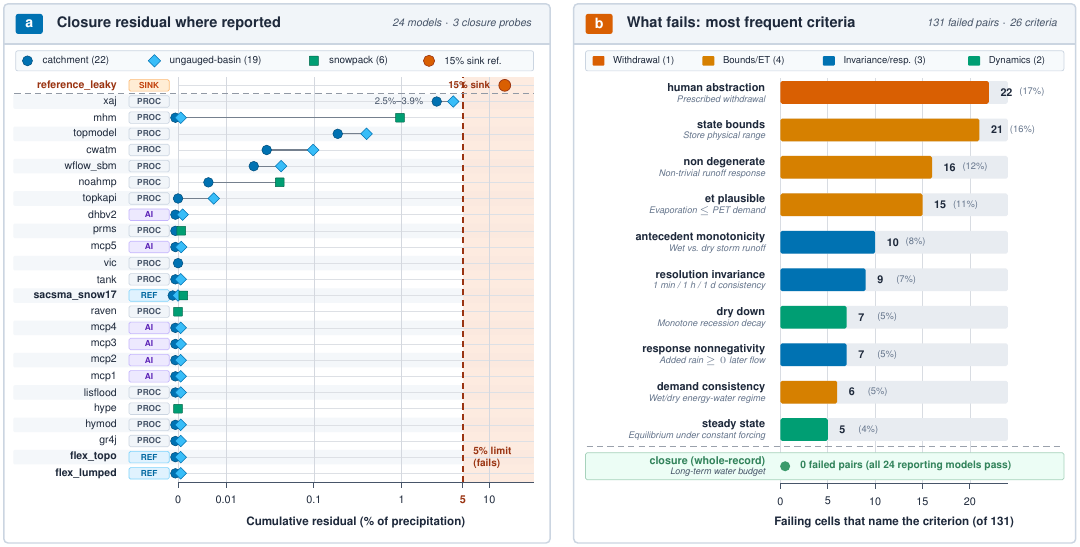}
\caption{Closure and failure modes in suite 0.1.0. \textbf{a}, Whole-record closure residuals that the evaluated models report on the closure probes (catchment, ungauged-basin and snowpack), as a share of cumulative precipitation, with 1 dot per model and probe, and the same quantity for the broken reference \code{reference\_leaky}. The dashed line marks the 5\% limit that the 4 closure probes share. \textbf{b}, Number of failed model-probe pairs in which each criterion was tripped. A pair can trip more than 1 criterion. The 131 failed pairs trip 26 distinct criteria, and the 10 most frequent are shown.}\label{fig:closure}
\end{figure}

The 131 failed pairs trip 26 distinct criteria (\cref{fig:closure}b). The 5 most frequent are \code{human\_abstraction} (22 pairs), \code{state\_bounds} (21), \code{non\_degenerate} (16), \code{et\_plausible} (15) and \code{antecedent\_monotonicity} (10). Four observations stand out.

First, no evaluated model trips the whole-record closure criterion on any of the 4 closure probes. Among the 22 models that report a residual on \code{catchment-closure}, the largest is 2.5\% of precipitation. Across the closure probes that report a residual, the largest is 3.9\% (\code{xaj} on \code{ungauged-basin-closure}; \cref{fig:closure}a). Closure of the long-term water budget holds across the evaluated models.

Second, a closing budget does not imply physical stores and fluxes. Five of 27 scored models fail \code{catchment-closure} on another criterion. The model \code{hype} lets soil storage reach 363~mm in a 320~mm profile. The models \code{raven} and \code{shud} evaporate 1.01 and 1.27 times the potential evaporation that the probe supplied. The model \code{simhyd} exceeds both its canopy capacity and the potential evaporation, and \code{summa} lets canopy storage reach 20.3~mm against a capacity of 2~mm. Although these models close their overall water budgets, they report states or fluxes that exceed physical capacity or atmospheric demand.

Third, closure over the record does not imply closure event by event. Four models fail the event-wise criterion of \code{mass/\allowbreak extreme-event-closure}. The model \code{xaj} fails on 133 of 1{,}316 events and \code{shud} on 130 of 1{,}257. The model \code{cwatm} fails on 4 of 1{,}257 events and \code{wflow\_sbm} on 1 of 1{,}299. The model \code{xaj} passes the whole-record criterion with a residual of 2.5\%, so a total over the record can hide events that lose or gain water. Three further models, \code{hype}, \code{simhyd} and \code{summa}, fail the same probe on state bounds.

Fourth, most models do not account for a prescribed human withdrawal. On \code{mass/\allowbreak human-abstraction}, 20 of 27 scored models show no change in any budget term when the probe withdraws 380~mm of water over 10 years. Two more account for part of it: \code{lisflood} leaves 83\% of the withdrawal unaccounted for, and \code{prms} leaves 29\%. Only 5 scored models close this budget within the 5\% limit, of which 3 are the exact references.

\subsection{Case study: an LSTM}\label{sec:case}
An LSTM flood-forecasting model illustrates how \HT{} evaluates a model that reports streamflow alone. We ran the publicly released weights of the model lineage behind \citet{nearing2024}, with the training forcing products mapped from the probe's precipitation, temperature and potential evaporation. Inputs that a probe cannot supply, such as net radiation and surface pressure, are derived from the forcing alone. Catchment attributes are taken from the first year of the generated record or set to the training mean. The model therefore operates as an average catchment under mock inputs. Our evaluation reflects the released weights coupled with this adapter rather than the full operational forecasting system.

The model reports runoff and discharge only. It is N/A on 26 of 37 probes. On 24 of them the probe requires evaporation, storage or heat fluxes, while on 2 the model and the probe disagree on an input or a time step. Rather than violating conservation on those 26 probes, the model cannot be falsified on them because it omits the required outputs. On the 11 probes that require runoff alone, it passes 5 and fails 6 (\cref{tab:lstm}).

\begin{table}[!tbp]
\centering\small
\caption{Verdicts of the LSTM on the 11 probes that it can be scored on, with the measured quantity on the seed that the report shows. Reference values are from the exact bucket on the same probe where stated.}\label{tab:lstm}
\begin{tabular}{@{}lcP{0.58\linewidth}@{}}
\toprule
Probe & Verdict & Measured quantity \\ \midrule
\code{resolution-invariance} & FAIL & Runoff differs by 73.8\% of precipitation between hourly and daily records of the same month (limit 10\%) \\
\code{extreme-rain} & FAIL & Across the ladder up to 10 times the largest storm, 35 of 526~mm of added rain run off (0.07). The exact bucket returns 0.56 to 1.00 \\
\code{dry-down} & FAIL & 2 rainless years yield a near-constant 0.3~mm\,d$^{-1}$ that does not decay (coefficient of variation 0.02, minimum 0.1) \\
\code{warming-response} & FAIL & Runoff falls by 0.088 per unit of added demand (limit 0.10). The exact bucket falls by 0.20 \\
\code{phase-counterfactual} & FAIL & Runoff differs by $-8.4$\% of the rain when snow falls as rain (limit 5\%) \\
\code{response-nonnegativity} & FAIL & Runoff falls 0.18~mm\,d$^{-1}$ below the control after 120~mm of rain is added \\
\code{causality} & PASS & Nothing changes before the added storm. Runoff responds with 0.32 of the added rain \\
\code{steady-state} & PASS & Settles to 0.63~mm\,d$^{-1}$ under 2.5~mm\,d$^{-1}$ of rain, with no variation in the last year \\
\code{runoff-bounds} & PASS & Runoff is 0.34 of the rain, inside the bounds 0.12 to 1.05 \\
\code{antecedent-monotonicity} & PASS & The wetter catchment runs off 16.4~mm more from a 60~mm storm (0.27 of it) \\
\code{area-invariance} & PASS & Scaling the catchment area leaves the depths unchanged \\ \bottomrule
\end{tabular}
\end{table}

The pattern is consistent: the model passes probes whose answer follows from its causal recurrent structure or a plausible climatological response, while failing every probe that moves the input outside the training regime. The passes are straightforward. The model uses no future information, settles under constant weather, keeps runoff within the bounds of precipitation, and produces more runoff on a wetter catchment. The failures fall into 3 groups. The first group concerns input extremes. Across the ladder up to 10 times the largest storm, the model returns only 7\% of the added rain, even though a catchment with a 320~mm soil capacity cannot absorb 526~mm. Over 2 rainless years, it maintains a baseflow floor learned from its training basins, so the discharge never recedes. The second group concerns perturbation responses. Warming produces the expected sign in both directions on every seed, but its magnitude falls below the 0.10 threshold on 3 of 5 seeds. Allowing the static climate attributes to update with the warmed record yields a stronger response, indicating that part of the apparent climate sensitivity is encoded in the static attributes rather than the dynamic weather forcing. The third group concerns physical consistency. Hourly records produce a runoff volume that differs by 73.8\% of precipitation from daily records of the same weather, because the network accepts arbitrary step sequences without an explicit time-step parameter and treats each hourly rate as a daily depth. Added rain also reduces subsequent daily runoff by 0.18~mm\,d$^{-1}$.

\subsection{Differentiable and mass-conserving models}
Two other AI-based families provide a contrast. The differentiable model \code{dhbv2} \citep{song2025} embeds an explicit HBV core, while the 5 mass-conserving perceptron checkpoints \code{mcp1} to \code{mcp5} \citep{wang2024mcp} route water through explicit stores with conservation built into the architecture.

The mass-conserving perceptron closes the water budget by construction, and the probes confirm this property. All 5 checkpoints pass \code{catchment-closure}, \code{extreme-event-closure} and \code{ungauged-basin-closure}, with whole-record residuals below 0.001\% of precipitation. However, they fail between 2 and 6 other probes, and the failed probes vary across training epochs within the same architecture. The checkpoint at epoch 27 (\code{mcp1}) fails 2 probes, while the checkpoint at epoch 2707 (\code{mcp4}) fails 6. In \code{mcp4}, runoff rises by 1.17 times the rain added when a storm is scaled by 2, falls below the control after rain is added, and still varies by 11.7\% (runoff) and 4.2\% (storage) in the final year of constant weather. Four checkpoints fail \code{antecedent-monotonicity}: the wetter catchment returns only 0.1 to 0.4~mm more runoff from a 60~mm storm in 3 checkpoints (below the 1.2~mm minimum), and 2.3~mm less in \code{mcp4}. Enforcing conservation in the architecture therefore secures that single law while leaving other physical behaviors dependent on training.

The differentiable model passes the 3 closure probes that it is scored on, together with the extreme-rain and warming probes. However, it fails 6 probes in the invariance and response families. Its runoff differs by 12.1\% of precipitation between hourly and daily records (limit 10\%), likely because the calibrated parameters retain the daily time-step scaling at which they were fitted. Runoff also differs by $-34.8$\% of the rain when snow falls as rain (limit 5\%). Added rain lowers subsequent runoff by 0.28~mm\,d$^{-1}$, and the precipitation-counterfactual probe records 10{,}040~mm of runoff in the control run against 8{,}686~mm of rain. Finally, the model fails the area-invariance probe because its channel store departs from the expected scaling when catchment area is rescaled, suggesting that channel parameters calibrated for 1 basin area do not transfer across scales. An explicit physical core does not by itself guarantee the invariances tested by the suite.

\section{Community protocol and governance}\label{sec:community}

A benchmark for physical laws needs more physics beyond what an individual group can supply. Mass, energy and momentum each have specialists who disagree about tolerances, denominators and what counts as a violation. We therefore built \HT{} as a protocol with explicit steps for contributing, reviewing and crediting work, and we describe it here because the 37 probes in this article exist through it.

\subsection{Contributing a probe or a model}
There are 2 routes. A contributor who wants to add a physical test claims an item from the public roadmap, or proposes a new probe, through an issue. The maintainer labels the proposal as accepted when the probe would discriminate between reference models, and the issue is assigned to the person who will do the work. This early review filters out weak probe designs before contributors invest time in implementation. The contributor then scaffolds the probe from a template with \code{ht init-probe}, writes the generator, the criteria and the reference models, and runs the acceptance gate. A contributor who wants a model evaluated proposes it through a separate form. The proposer does not need to package the model. The issue is assigned to whoever will build the container, which is often the maintainer.

Both routes are written for coding agents as well as for people. The repository contains an \code{AGENTS.md} file that states the input-output contract, the lexicon and the acceptance rules in compact form. An agent given a model repository and this file can draft an adapter, and the contract review below then checks it. Because a failed probe is a scientific finding rather than a packaging defect, physical failures are never grounds for rejecting a model.

\subsection{Review and decisions}
Every probe merged from 16 September 2026 is reviewed by at least 2 members of a reviewer pool for its law. The probe is merged only after both reviewers approve and the gate is green. The pool currently has 3 reviewers for energy and 3 for mass. Probes merged before 16 September were accepted by the maintainer under the earlier process. Each model undergoes a single review focused on contract compliance rather than physical performance: the adapter must honor the input-output directory, the declared outputs must be honest, and the image must be reproducible.

A scientific steering committee of 4 to 6 people, spanning hydrologic modeling, machine learning for Earth science and operational hydrology, is being formed. It will settle disputes about probe design, decide when a suite version is cut, and approve any change to the tolerance of a merged probe together with the authors of that probe. Such a change silently alters every recorded result, so it bumps the minor version of the suite.

The repository lists 25 contributors beyond the maintainer, namely 18 authors of merged probes and 7 further proposers of accepted models. The 18 probe authors and 1 proposer with 5 accepted proposals are authors of this article. The remaining 6 proposers are credited in the contributor file. 

\subsection{Growth and roadmap}
\begin{figure}[!tbp]
\centering
\includegraphics[width=\linewidth]{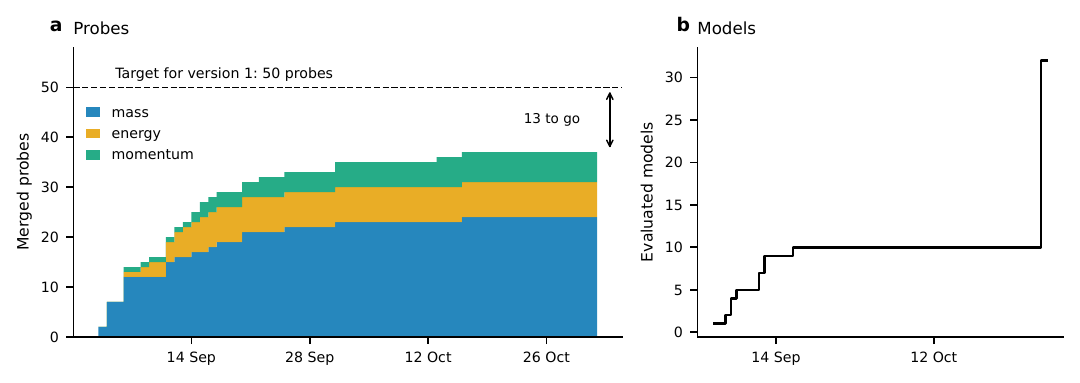}
\caption{Growth of \HT{} from its first probe to suite 0.1.0. Dates are those of the commits that merged each probe and each model into the main branch. The dashed line marks the target of 50 probes for version~1. Of the 32 models, 22 were added on 31 October 2026.}\label{fig:growth}
\end{figure}

The suite grew from 1 probe on 3 September 2026 to 37 on 16 October (\cref{fig:growth}). The model pool grew from 1 to 10 by 17 September and to 32 on 31 October. The roadmap lists the probes that we want next, with the template to start from and a note on what each probe discriminates. Two further probes are claimed and in progress, a network-wide mass closure probe and a snowpack ripening probe. One probe remains unclaimed, the reach-by-reach mass balance of a channel. We aim for 50 probes at version~1, with 13 to add, and for more energy and momentum probes in particular.

\section{Discussion and limitations}\label{sec:discussion}

\subsection{What the verdicts show}
Three results stand out. First, evaluated models rarely fail on long-term water-budget closure. All 22 models that report a residual on \code{catchment-closure} stay below 2.6\% of precipitation; instead, failures occur in the response of the model to a perturbed input and in the physical bounds of reported states and fluxes. Second, enforcing mass conservation in a network architecture secures that law alone without guaranteeing physical behavior on other laws or regimes. Although the 5 mass-conserving perceptron checkpoints close the water budget within 0.001\%, they still fail between 2 and 6 probes each. Third, nearly half of all model-probe pairs (49\%) cannot be scored, primarily because models omit the variables that a probe requires. Reporting internal states and fluxes is therefore the most direct way to make a model falsifiable.

\subsection{Use before deployment and during training}
We see 2 use cases of \HT{}. A modeler can run \HT{} as a smoke test before a model enters operations, in the way a unit test suite precedes a software release. The verdicts name the violated law and give the measured size of the violation. A forecaster or a water manager needs this information to decide whether a model can support drought assessment, irrigation planning or a climate impact study that relies on evaporation and storage. Journals can also run submitted models through the suite as a reproducibility and consistency check, because the suite is open and every verdict can be reproduced from a recorded seed.

The probes can also serve during training. The mass-conserving perceptron checkpoints show that verdicts change from epoch to epoch within 1 architecture. The checkpoint at epoch 27 fails 2 probes, while the checkpoint at epoch 2707 fails 6. A model developer can therefore run the probes on saved checkpoints and select among them based on physical consistency as well as hydrograph skill. Although the probes are black-box experiments and cannot enter a gradient-based loss function directly, they identify which physical constraints a training objective should add.

\subsection{Limitations}
First, passing the benchmark is a necessary condition for physical consistency rather than a sufficient condition for predictive accuracy. A model that passes every probe conserves mass, energy and momentum under the tested conditions, but it may still be inaccurate on an observed basin or violate a law under untested conditions. The suite therefore complements rather than replaces hydrograph evaluation.

Second, public probes can be targeted. A fresh seed prevents a model from memorizing a case, but it does not prevent tuning to the rule that a probe tests. The suite offers 2 defenses. A private probe tree can be added with \code{--probe-root}, so that an evaluator can hold out probes whose generators and tolerances are unpublished. Each verdict also reports the measured quantity, so that a model that barely passes can be identified. We therefore regard verdicts from the public suite as a pre-deployment smoke test rather than a certification.

Third, tolerances are conventions. The 5\% closure limit, the 10\% resolution limit and the other thresholds are engineering choices, and the sweep in \cref{sec:refs} shows that the closure verdicts of the evaluated models do not depend on the exact value between 2.6\% and 15\%. For other criteria, the gate ensures that exact references pass and broken references fail. Changes to a tolerance require the governance process in \cref{sec:community}.

Fourth, the catchment and the weather are synthetic. Most probes run on 1 invented lumped catchment of 250~km$^2$ with a 320~mm soil and a 2~mm canopy, under weather from a generator whose climate does not match any gauged basin. Only the ungauged-basin probe samples other attributes. This design is a strength for controlled experiments because the exact physical response is known. However, a physical violation and a distribution shift cannot always be separated. A model trained on real basins that misbehaves in the synthetic catchment may violate physics or may operate far outside its training distribution. The adapter adds a second dependence. In an early evaluation of the differentiable model, 1 reasonable choice for a seasonality attribute changed the coefficient of variation of runoff from 1.1 to 0.07. The probe does not provide this attribute, and that choice changed 2 verdicts. A verdict therefore evaluates a model together with its adapter, and the adapter is documented alongside each model. We plan a real-data track after version~1 is published. It will keep the weather synthetic while replacing the invented catchment with gauged basins whose attributes are drawn from Caravan \citep{kratzert2023}. The repository holds a proposal for 4 CAMELS-US basins \citep{addor2017} that span an alpine snowpack, a maritime rain regime, a karst aquifer and a lowland agricultural catchment.

Fifth, the spatial scale and scope remain limited. The probes treat a catchment as 1 lumped control volume, apart from the routing and hydraulics probes on a single reach or a small network. They do not test spatial patterns within a basin, even though semi-distributed structure can improve streamflow prediction \citep{boyle2001}. Momentum is the least developed law, and no probe yet tests the full unsteady momentum balance. The energy results also rest on a small sample of models: only 3 evaluated models are scored on the 2 daily cross-budget probes, and all 3 fail them. The repository documentation notes that daily averaging does not remove the mismatch between the net radiation that these probes supply and the radiation that an adapter computes internally. An open issue tracks this fix, and we regard those archived scores as provisional.

Sixth, a black-box test verifies the internal consistency of the variables a model reports rather than the physical realism of its internal states. A model that reports a synthetic store can close its budget on paper. The paired and counterfactual probes make such a post-hoc adjustment harder to sustain because the reported store must respond physically to perturbations, but they cannot rule it out completely.

Seventh, each model enters with 1 version and 1 adapter. Verdicts for stochastic models depend on the random draw unless the adapter fixes it, as the LSTM adapter does by taking the exact median of the predictive distribution. Scores are comparable only within 1 suite version.

\section{Conclusion}\label{sec:conclusion}

A hydrograph score cannot distinguish a model that conserves water from one that loses 15\% of it, yet the models that issue operational forecasts are judged mainly by that score. We introduced \HT, a benchmark of 37 physics probes for the conservation of mass, energy and momentum in hydrologic models. First, each probe is a controlled experiment on forcing generated at run time, with binary criteria on 1 physical law. Second, the harness runs the model as a black box and keeps the labels and tolerances on the host. Third, every probe is validated against 77 reference models, including 18 that must pass and 59 that must fail. Across 32 evaluated models, only the 3 exact references pass every probe that they are scored on among the 28 models scored on 5 or more probes. A pre-trained flood-forecasting LSTM fails 6 of the 11 probes it can be scored on, while 49\% of all model-probe pairs remain unscored because models do not report the variables that the probes require.

These findings support 3 applications. Model developers can run the suite as a pre-deployment smoke test and expose additional internal states and fluxes. Journal editors can verify physical consistency alongside reported hydrograph metrics, and operational agencies can require benchmark verdicts before adopting a model. Under our open community protocol, authors of merged probes become co-authors of the benchmark. We invite the community to contribute the 13 probes needed to reach the 50-probe target for version~1, to propose additional models, and to help build the upcoming real-catchment evaluation track.

\section*{Code and data availability}
The harness, the 37 probes, the 77 reference models, the model adapters and the results archive are available at \url{https://github.com/Flood-Lab/HydroTuring}, and the project website is \url{https://flood-lab.github.io/HydroTuring/}. Suite 0.1.0 \citep{hydroturing2026} is archived on Zenodo (\url{https://doi.org/10.5281/zenodo.23874854}), and all versions share the concept DOI \url{https://doi.org/10.5281/zenodo.23874853}. The harness, probes, reference models and adapters are released under the Creative Commons Attribution-NonCommercial 4.0 license, and each packaged model keeps its own license, which its manifest names. No dataset is distributed, because every probe generates its forcing from a recorded seed. The verdicts, the measured quantities and the seeds of every run reported here are in the results archive in the repository.

\section*{Acknowledgments}
We thank the contributors who proposed models, reviewed probes and reported issues in the repository.

\bibliographystyle{plainnat}
\bibliography{references}

\appendix
\section{Schemas and a worked probe}\label{app:lexicon}

\subsection{Model manifest}
A model declares what it emits and what it needs in \code{model.yaml}. The manifest of the exact bucket illustrates the fields. A model that predicts discharge only declares \code{fluxes: [mrro, dis]} and \code{states: []}, and it is then N/A (INCOMPLETE) on every probe that needs more.

{\scriptsize
\begin{verbatim}
name: reference_bucket
version: "1.1.0"
license: CC-BY-NC-4.0
runner: subprocess
entrypoint: ["python3", "ht_adapter.py"]
timestep: [PT1D, PT1H, PT15M, PT5M, PT1M]
emits:
  fluxes: [pr, snm, evspsbl, mrro, dis, gwex]
  states: [mrso, snw, canopy, channel]
  diagnostics: [stage]
needs_forcing: [pr, tas, pet]
supports:
  perturbation: true
needs_static: [soil_capacity_mm, canopy_capacity_mm]
uses_static: [area_km2, width_m, cross_section_shape, bed_elevation_m,
              slope, manning_n]
\end{verbatim}
}

\subsection{Probe definition}
A probe is a directory with a definition file, a generator and a README. The definition of the reference probe, \code{mass/\allowbreak catchment-closure}, is shown in abridged form. The \code{baselines} block is the declaration that the acceptance gate enforces.

{\scriptsize
\begin{verbatim}
id: mass/catchment-closure
law: mass
requires:
  fluxes: [pr, evspsbl, mrro]
  states: [mrso, snw, canopy]
case:
  generator: generate.py      # forcing is generated from the seed at run time
  n_seeds: 5
  timestep: PT1D
  period_years: 10
  spinup_days: 365
  max_runtime_s: 60
criteria:
  - closure:          {denominator: sum_pr, threshold: 0.05,
                       sinks: [evspsbl, mrro]}
  - state_bounds:     {mrso: [0.0, soil_capacity_mm], snw: [0.0, null],
                       canopy: [0.0, canopy_capacity_mm]}
  - et_plausible:     {max_ratio_to_pet: 1.0, pet_variable: pet}
  - non_degenerate:   {runoff_ratio: [0.02, 0.98], min_flux_cv: 0.1,
                       min_response: 0.05, cv_variables: [mrro, evspsbl]}
  - forcing_fidelity: {variables: [pr], rtol: 1.0e-6}
baselines:
  must_pass: [reference_bucket, flex_lumped, flex_topo, sacsma_snow17]
  must_fail:
    reference_leaky: closure
    reference_cheater: state_bounds
    reference_degenerate: non_degenerate
\end{verbatim}
}

\subsection{Lexicon}
\Cref{tab:lexfull} lists the output variables of the lexicon, with their group, units and meaning. The forcing variables and static attributes are defined in the same file of the repository.

{\small\begin{longtable}{@{}P{0.16\linewidth}P{0.1\linewidth}P{0.1\linewidth}P{0.55\linewidth}@{}}
\caption{Full lexicon of the \texttt{/io} contract (schemas/lexicon.yaml, suite 0.1.0). Over a row of several model steps, a variable is a mean, an end-of-row value, or an instantaneous value, as stated in the repository.}\label{tab:lexfull}\\
\toprule Name & Group & Units & Physical meaning \\ \midrule \endfirsthead
\toprule Name & Group & Units & Physical meaning \\ \midrule \endhead
\bottomrule \endfoot
\texttt{pr} & flux & mm d$^{-1}$ & Precipitation, echoed exactly as the forcing gave it \\
\texttt{evspsbl} & flux & mm d$^{-1}$ & Evapotranspiration, including sublimation; net of dew and frost where the model nets them \\
\texttt{mrro} & flux & mm d$^{-1}$ & Total runoff \\
\texttt{dis} & flux & m$^3$ s$^{-1}$ & River discharge \\
\texttt{gwex} & flux & mm d$^{-1}$ & Declared exchange with the outside: regional groundwater, inter-basin transfer \\
\texttt{gw\_\allowbreak{}sw\_\allowbreak{}exchange} & flux & mm d$^{-1}$ & Net river-aquifer exchange inside the control volume, not added to \texttt{gwex} or counted in closure \\
\texttt{gw\_\allowbreak{}to\_\allowbreak{}sw} & flux & mm d$^{-1}$ & The aquifer losing to the river, a component of \texttt{gw\_sw\_exchange} \\
\texttt{sw\_\allowbreak{}to\_\allowbreak{}gw} & flux & mm d$^{-1}$ & The river losing to the aquifer, a component of \texttt{gw\_sw\_exchange} \\
\texttt{gw\_\allowbreak{}boundary} & flux & mm d$^{-1}$ & Every other flux across the aquifer's own boundary, such as a GHB or a well \\
\texttt{sbl} & flux & mm d$^{-1}$ & The sublimating share of \texttt{evspsbl}, a component of it rather than an addition \\
\texttt{snm} & flux & mm d$^{-1}$ & Liquid delivered from the snow module to the ground, rain passed through included, and not an outflow beside \texttt{mrro} \\
\texttt{hfls} & flux & W m$^{-2}$ & Latent heat flux \\
\texttt{hfss} & flux & W m$^{-2}$ & Sensible heat flux \\
\texttt{hfg} & flux & W m$^{-2}$ & Ground heat flux at the actual soil surface, where a deeper flux is corrected for the heat stored above it \\
\texttt{rlus} & flux & W m$^{-2}$ & Total upward longwave: emission plus reflected downward longwave \\
\texttt{hfg\_\allowbreak{}bottom} & flux & W m$^{-2}$ & Heat flux through the bottom of the specified soil layer \\
\texttt{mrso} & state & mm & Soil water storage (liquid and ice), reported as an absolute state rather than a tendency \\
\texttt{snw} & state & mm & Snow water equivalent, identically zero and named in \texttt{zero\_stores} for a model with no snow module \\
\texttt{canopy} & state & mm & Canopy interception storage, identically zero and named in \texttt{zero\_stores} for a model with no such store \\
\texttt{gw} & state & mm & Groundwater storage below the soil column \\
\texttt{channel} & state & mm & Water generated as runoff but not yet released by the model's routing \\
\texttt{ts} & diagnostic & K & Radiative surface (skin) temperature \\
\texttt{tsoil\_\allowbreak{}layer} & diagnostic & K & Mean temperature of the specified soil layer \\
\texttt{stage} & diagnostic & m & Water-surface elevation on the case's datum, read beside the row's mean discharge (depth is \texttt{stage} minus \texttt{bed\_elevation\_m}) \\
\texttt{lwsnl} & diagnostic & mm & Every non-solid component of \texttt{snw}, so \texttt{snw} minus \texttt{lwsnl} is the ice (part of \texttt{snw}, not additional) \\
\texttt{csnow} & diagnostic & J m$^{-2}$ & The pack's cold content: the positive energy still needed to bring its ice to $0\,^\circ$C \\
\texttt{reach\_\allowbreak{}q\_\allowbreak{}in} & routing & m$^3$ s$^{-1}$ & Total inflow into the identified reach, in \texttt{routing.csv} \\
\texttt{q\_\allowbreak{}out} & routing & m$^3$ s$^{-1}$ & Flow across the reach's downstream boundary, in \texttt{routing.csv} \\
\texttt{channel\_\allowbreak{}storage} & routing & m$^3$ & Absolute water volume in the reach's routed channel, in \texttt{routing.csv} \\
\midrule
\texttt{pr} & forcing & mm d$^{-1}$ & Precipitation \\
\texttt{tas} & forcing & $^\circ$C & Near-surface air temperature in Celsius, unlike the kelvin of \texttt{ts} and \texttt{tsoil\_layer} \\
\texttt{pet} & forcing & mm d$^{-1}$ & Potential evapotranspiration \\
\texttt{rn} & forcing & W m$^{-2}$ & Prescribed net radiation, which the model accepts as the energy input to the reported budget when declared \\
\texttt{rsds} & forcing & W m$^{-2}$ & Downward shortwave radiation \\
\texttt{rlds} & forcing & W m$^{-2}$ & Downward longwave radiation \\
\texttt{huss} & forcing & kg kg$^{-1}$ & Near-surface specific humidity \\
\texttt{ps} & forcing & Pa & Surface air pressure \\
\texttt{sfcWind} & forcing & m s$^{-1}$ & Near-surface wind speed \\
\texttt{abstr} & forcing & mm d$^{-1}$ & Prescribed human withdrawal, net of return flow \\
\texttt{gwh} & forcing & m & External hydraulic head behind the model's \texttt{gwex}; declaring it asserts \texttt{gwex} responds monotonically to it \\
\texttt{gw\_\allowbreak{}recharge} & forcing & mm d$^{-1}$ & Direct recharge to the aquifer \\
\texttt{sw\_\allowbreak{}stage\_\allowbreak{}m} & forcing & m & River stage the aquifer exchanges with \\
\texttt{q\_\allowbreak{}in} & forcing & m$^3$ s$^{-1}$ & Prescribed river or upstream inflow, already a discharge and not multiplied by area or row length \\
\midrule
\texttt{area\_\allowbreak{}km2} & static & km$^2$ & Catchment area the fluxes are converted over \\
\texttt{elevation\_\allowbreak{}m} & static & m & Catchment elevation \\
\texttt{latitude\_\allowbreak{}deg} & static & degrees north & Latitude \\
\texttt{main\_\allowbreak{}channel\_\allowbreak{}length\_\allowbreak{}km} & static & km & Length of the longest channel \\
\texttt{centroid\_\allowbreak{}channel\_\allowbreak{}length\_\allowbreak{}km} & static & km & Channel length from the outlet to the point nearest the catchment's centroid \\
\texttt{soil\_\allowbreak{}capacity\_\allowbreak{}mm} & static & mm & Water the soil column can hold \\
\texttt{canopy\_\allowbreak{}capacity\_\allowbreak{}mm} & static & mm & Water the canopy can intercept (0 is a bare surface) \\
\texttt{degree\_\allowbreak{}day\_\allowbreak{}factor\_\allowbreak{}mm\_\allowbreak{}per\_\allowbreak{}C\_\allowbreak{}day} & static & mm $^\circ$C$^{-1}$ d$^{-1}$ & Snowmelt per degree-day \\
\texttt{baseflow\_\allowbreak{}coefficient} & static & d$^{-1}$ & Fraction of soil water drained to baseflow per day \\
\texttt{snow\_\allowbreak{}threshold\_\allowbreak{}degC} & static & $^\circ$C & Air temperature dividing snow from rain \\
\texttt{eps} & static & 1 & Surface emissivity \\
\texttt{soil\_\allowbreak{}layer\_\allowbreak{}depth\_\allowbreak{}m} & static & m & Depth of the prescribed soil layer, from the surface \\
\texttt{soil\_\allowbreak{}heat\_\allowbreak{}capacity\_\allowbreak{}areal} & static & J m$^{-2}$ K$^{-1}$ & Effective areal heat capacity of that layer, already including the depth \\
\texttt{soil\_\allowbreak{}temperature\_\allowbreak{}initial} & static & K & Initial temperature of that layer \\
\texttt{soil\_\allowbreak{}solid\_\allowbreak{}heat\_\allowbreak{}capacity} & static & J m$^{-3}$ K$^{-1}$ & Volumetric heat capacity of the soil solids \\
\texttt{soil\_\allowbreak{}water\_\allowbreak{}heat\_\allowbreak{}capacity} & static & J m$^{-3}$ K$^{-1}$ & Volumetric heat capacity of soil water \\
\texttt{soil\_\allowbreak{}pore\_\allowbreak{}air\_\allowbreak{}heat\_\allowbreak{}capacity} & static & J m$^{-3}$ K$^{-1}$ & Volumetric heat capacity of the air in the pores \\
\texttt{soil\_\allowbreak{}porosity} & static & 1 & Volumetric porosity \\
\texttt{soil\_\allowbreak{}water\_\allowbreak{}content\_\allowbreak{}initial} & static & m$^3$ m$^{-3}$ & Initial volumetric water content \\
\texttt{aquifer\_\allowbreak{}top\_\allowbreak{}m} & static & m & Elevation of the aquifer's top \\
\texttt{aquifer\_\allowbreak{}bottom\_\allowbreak{}m} & static & m & Elevation of the aquifer's bottom \\
\texttt{aquifer\_\allowbreak{}initial\_\allowbreak{}head\_\allowbreak{}m} & static & m & Initial aquifer head \\
\texttt{aquifer\_\allowbreak{}specific\_\allowbreak{}yield} & static & 1 & Drainable porosity above the water table \\
\texttt{aquifer\_\allowbreak{}storage\_\allowbreak{}coefficient} & static & 1 & Storativity, not specific storage \\
\texttt{river\_\allowbreak{}conductance\_\allowbreak{}m2\_\allowbreak{}per\_\allowbreak{}day} & static & m$^2$ d$^{-1}$ & Conductance between the river and the aquifer \\
\texttt{river\_\allowbreak{}bottom\_\allowbreak{}offset\_\allowbreak{}m} & static & m & River stage minus riverbed elevation \\
\texttt{routing\_\allowbreak{}network} & static & - & Object whose \texttt{reaches} member lists the unique string IDs of the reaches \\
\texttt{reach\_\allowbreak{}length\_\allowbreak{}m} & static & m & Length of the reach \\
\texttt{width\_\allowbreak{}m} & static & m & Width of the rectangular channel bed \\
\texttt{bankfull\_\allowbreak{}depth\_\allowbreak{}m} & static & m & Depth at which the channel is full \\
\texttt{bed\_\allowbreak{}elevation\_\allowbreak{}m} & static & m & Elevation of the rigid bed on the datum stage uses \\
\texttt{slope} & static & 1 & Bed slope \\
\texttt{manning\_\allowbreak{}n} & static & s m$^{-1/3}$ & Manning's roughness \\
\texttt{cross\_\allowbreak{}section\_\allowbreak{}shape} & static & - & Name of the cross-section's shape, such as rectangular \\
\end{longtable}}

\section{Probe catalog}\label{app:catalogue}
\Cref{tab:catalogue} lists the 37 probes of suite 0.1.0 by law and family. \Cref{tab:authors} gives the authors of every probe as recorded in its definition file. \Cref{tab:proposers} gives the proposer of every accepted model.

{\footnotesize
\begin{longtable}{@{}P{0.22\linewidth}P{0.38\linewidth}P{0.26\linewidth}c@{}}
\caption{Probe catalog for suite version 0.1.0. Criteria shared by most probes (\texttt{state\_bounds}, \texttt{non\_degenerate}, \texttt{forcing\_fidelity}) are listed only when they are the sole criteria. Seeds are the number of generated cases per probe.}\label{tab:catalogue}\\
\toprule Probe & Question the probe asks & Scored criteria & Seeds \\ \midrule \endfirsthead
\toprule Probe & Question the probe asks & Scored criteria & Seeds \\ \midrule \endhead
\bottomrule \endfoot
\multicolumn{4}{@{}l}{\textit{Mass: closure}} \\
\texttt{catchment-\allowbreak{}closure} & Catchment water balance closure under synthetic forcing & \texttt{closure}, \texttt{et\_\allowbreak{}plausible} & 5 \\
\texttt{snowpack-\allowbreak{}mass-\allowbreak{}closure} & Internal snowpack water-balance closure & \texttt{closure}, \texttt{snowpack\_\allowbreak{}response} & 5 \\
\texttt{extreme-\allowbreak{}event-\allowbreak{}closure} & Event-wise water balance closure under extreme rainfall & \texttt{closure}, \texttt{event\_\allowbreak{}water\_\allowbreak{}closure} & 5 \\
\texttt{ungauged-\allowbreak{}basin-\allowbreak{}closure} & Water balance integrity across declared catchment attribute domains & \texttt{closure}, \texttt{et\_\allowbreak{}plausible} & 12 \\
\multicolumn{4}{@{}l}{\textit{Mass: invariance}} \\
\texttt{resolution-\allowbreak{}invariance} & Integrated runoff must not depend on the step the weather was given at & \texttt{resolution\_\allowbreak{}invariance} & 3 \\
\texttt{area-\allowbreak{}invariance} & The same weather on a 10 times larger catchment gives the same depths & \texttt{invariance} & 3 \\
\texttt{time-\allowbreak{}origin-\allowbreak{}invariance} & Identical forcing under aligned calendars gives identical water budgets & \texttt{invariance}, \texttt{closure} & 3 \\
\texttt{groundwater-\allowbreak{}datum-\allowbreak{}invariance} & Groundwater exchange is invariant to a common vertical datum shift & \texttt{datum\_\allowbreak{}flux\_\allowbreak{}invariance}, \texttt{exchange\_\allowbreak{}directions}, \texttt{groundwater\_\allowbreak{}balance}, \texttt{exchange\_\allowbreak{}components} & 3 \\
\texttt{spinup-\allowbreak{}cycle-\allowbreak{}invariance} & A periodic climate must forget extra spin-up cycles & \texttt{closure}, \texttt{spinup\_\allowbreak{}cycle\_\allowbreak{}invariance} & 3 \\
\multicolumn{4}{@{}l}{\textit{Mass: response}} \\
\texttt{warming-\allowbreak{}response} & Warmer air over the same rain must mean less runoff, and cooler air more & \texttt{response\_\allowbreak{}sign} & 5 \\
\texttt{precipitation-\allowbreak{}counterfactual} & Added or removed precipitation must be partitioned among the budget terms & \texttt{closure}, \texttt{counterfactual\_\allowbreak{}response}, \texttt{monotone\_\allowbreak{}response} & 3 \\
\texttt{phase-\allowbreak{}counterfactual} & Rain that would have been snow is still the same water & \texttt{phase\_\allowbreak{}invariance} & 3 \\
\texttt{response-\allowbreak{}nonnegativity} & More water can never lower the flow, on any day & \texttt{response\_\allowbreak{}nonnegativity} & 3 \\
\texttt{antecedent-\allowbreak{}monotonicity} & The same storm on a wetter catchment runs off more, and no more than the extra water & \texttt{antecedent\_\allowbreak{}monotonicity} & 3 \\
\texttt{extreme-\allowbreak{}rain} & More rain cannot mean less runoff, nor more runoff than was added & \texttt{monotone\_\allowbreak{}response} & 3 \\
\multicolumn{4}{@{}l}{\textit{Mass: dynamics}} \\
\texttt{causality} & A catchment cannot respond to rain that has not fallen yet & \texttt{causality} & 3 \\
\texttt{dry-\allowbreak{}down} & Without rain, runoff can only fall, and only stored water can drain & \texttt{dry\_\allowbreak{}down} & 3 \\
\texttt{steady-\allowbreak{}state} & Under constant weather the catchment settles and the budget balances & \texttt{steady\_\allowbreak{}state} & 3 \\
\texttt{runoff-\allowbreak{}bounds} & Runoff is bounded by rain minus demand below and by rain above & \texttt{runoff\_\allowbreak{}bounds} & 5 \\
\texttt{multi-\allowbreak{}decadal-\allowbreak{}drift} & Slow budget-compensating storage drift over repeated climate & \texttt{closure}, \texttt{total\_\allowbreak{}storage\_\allowbreak{}drift}, \texttt{et\_\allowbreak{}plausible} & 5 \\
\multicolumn{4}{@{}l}{\textit{Mass: groundwater and human}} \\
\texttt{aquifer-\allowbreak{}recharge-\allowbreak{}ordering} & Aquifer recharge preserves storage and exchange ordering & \texttt{aquifer\_\allowbreak{}recharge\_\allowbreak{}ordering}, \texttt{groundwater\_\allowbreak{}balance}, \texttt{exchange\_\allowbreak{}components} & 3 \\
\texttt{exchange-\allowbreak{}response} & A declared head-driven exchange must respond to its external head & \texttt{closure}, \texttt{exchange\_\allowbreak{}response} & 5 \\
\texttt{gw-\allowbreak{}sw-\allowbreak{}exchange-\allowbreak{}consistency} & Groundwater-river exchange and storage close the water budget & \texttt{exchange\_\allowbreak{}components}, \texttt{groundwater\_\allowbreak{}balance}, \texttt{exchange\_\allowbreak{}directions} & 5 \\
\texttt{human-\allowbreak{}abstraction} & Prescribed irrigation withdrawal must leave the budget & \texttt{closure}, \texttt{human\_\allowbreak{}abstraction} & 3 \\
\midrule \multicolumn{4}{@{}l}{\textit{Energy: single budget}} \\ \midrule
\texttt{pet-\allowbreak{}consistency} & Evaporation follows demand when the soil is wet and water when it is dry & \texttt{demand\_\allowbreak{}consistency} & 3 \\
\texttt{surface-\allowbreak{}energy-\allowbreak{}closure} & Hourly surface budgets cannot hide canceling energy errors & \texttt{energy\_\allowbreak{}closure\_\allowbreak{}by\_\allowbreak{}phase} & 5 \\
\texttt{soil-\allowbreak{}heat-\allowbreak{}storage-\allowbreak{}consistency} & Soil heat fluxes and layer temperature must agree & \texttt{soil\_\allowbreak{}heat\_\allowbreak{}storage} & 5 \\
\texttt{radiation-\allowbreak{}consistency} & Surface temperature and upward longwave must describe one surface & \texttt{radiative\_\allowbreak{}identity} & 5 \\
\multicolumn{4}{@{}l}{\textit{Energy: cross budget}} \\
\texttt{latent-\allowbreak{}heat-\allowbreak{}et-\allowbreak{}consistency} & Latent heat and evapotranspiration must describe one evaporation & \texttt{closure}, \texttt{energy\_\allowbreak{}closure}, \texttt{flux\_\allowbreak{}identity}, \texttt{et\_\allowbreak{}plausible} & 5 \\
\texttt{evaporative-\allowbreak{}partition} & Energy a drying surface stops evaporating must warm the air & \texttt{partition\_\allowbreak{}shift}, \texttt{closure}, \texttt{energy\_\allowbreak{}closure}, \texttt{flux\_\allowbreak{}identity} & 3 \\
\texttt{snowmelt-\allowbreak{}energy-\allowbreak{}water} & Melt reported as water must be paid for out of the energy budget & \texttt{closure}, \texttt{melt\_\allowbreak{}energy} & 5 \\
\midrule \multicolumn{4}{@{}l}{\textit{Momentum: routing}} \\ \midrule
\texttt{routing-\allowbreak{}conservation} & A routing store holds runoff for as long as its hydrograph and no longer & \texttt{routing\_\allowbreak{}conservation} & 3 \\
\texttt{routing-\allowbreak{}lag-\allowbreak{}consistency} & Runoff peak lag follows the catchment geomorphic travel-time scale & \texttt{lag\_\allowbreak{}time\_\allowbreak{}bounds}, \texttt{scaling\_\allowbreak{}monotonicity} & 3 \\
\multicolumn{4}{@{}l}{\textit{Momentum: hydraulics}} \\
\texttt{stage-\allowbreak{}discharge-\allowbreak{}monotonic} & A stage reading rises with discharge, and loops the way a flood wave does & \texttt{rating\_\allowbreak{}monotonic}, \texttt{rating\_\allowbreak{}loop} & 3 \\
\texttt{uniform-\allowbreak{}flow-\allowbreak{}friction-\allowbreak{}consistency} & Steady discharge and stage satisfy the declared open-channel friction balance & \texttt{uniform\_\allowbreak{}flow\_\allowbreak{}friction} & 5 \\
\texttt{froude-\allowbreak{}regime} & A mild-sloped reach cannot outrun its own wave & \texttt{froude\_\allowbreak{}subcritical} & 3 \\
\texttt{wave-\allowbreak{}celerity-\allowbreak{}bounds} & Transient flood-wave celerity responds to hydraulic state & \texttt{wave\_\allowbreak{}celerity\_\allowbreak{}bounds} & 3 \\
\end{longtable}}

{\footnotesize
\begin{longtable}{@{}P{0.3\linewidth}P{0.62\linewidth}@{}}
\caption{Authors of each merged probe, as recorded in \texttt{probe.yaml}.}\label{tab:authors}\\
\toprule Probe & Authors \\ \midrule \endfirsthead
\toprule Probe & Authors \\ \midrule \endhead
\bottomrule \endfoot
\texttt{mass/catchment-\allowbreak{}closure} & Zhi Li \\
\texttt{mass/snowpack-\allowbreak{}mass-\allowbreak{}closure} & Jinlong Hu \\
\texttt{mass/extreme-\allowbreak{}event-\allowbreak{}closure} & Taiqi Lian \\
\texttt{mass/ungauged-\allowbreak{}basin-\allowbreak{}closure} & Shunan Zhou \\
\texttt{mass/resolution-\allowbreak{}invariance} & Zhi Li \\
\texttt{mass/area-\allowbreak{}invariance} & Zhi Li \\
\texttt{mass/time-\allowbreak{}origin-\allowbreak{}invariance} & Siavash Shams \\
\texttt{mass/groundwater-\allowbreak{}datum-\allowbreak{}invariance} & Zhenjiang Wu \\
\texttt{mass/spinup-\allowbreak{}cycle-\allowbreak{}invariance} & Kaihao Long \\
\texttt{mass/warming-\allowbreak{}response} & Zhi Li \\
\texttt{mass/precipitation-\allowbreak{}counterfactual} & Qingyi Yang \\
\texttt{mass/phase-\allowbreak{}counterfactual} & Zhi Li \\
\texttt{mass/response-\allowbreak{}nonnegativity} & Zhi Li \\
\texttt{mass/antecedent-\allowbreak{}monotonicity} & Zhi Li \\
\texttt{mass/extreme-\allowbreak{}rain} & Zhi Li \\
\texttt{mass/causality} & Zhi Li \\
\texttt{mass/dry-\allowbreak{}down} & Zhi Li \\
\texttt{mass/steady-\allowbreak{}state} & Zhi Li \\
\texttt{mass/runoff-\allowbreak{}bounds} & Zhi Li \\
\texttt{mass/multi-\allowbreak{}decadal-\allowbreak{}drift} & Bing Li \\
\texttt{mass/aquifer-\allowbreak{}recharge-\allowbreak{}ordering} & Zhenjiang Wu \\
\texttt{mass/exchange-\allowbreak{}response} & Songkun Yan \\
\texttt{mass/gw-\allowbreak{}sw-\allowbreak{}exchange-\allowbreak{}consistency} & Yaji Wang \\
\texttt{mass/human-\allowbreak{}abstraction} & Yuanhang Liu \\
\texttt{energy/pet-\allowbreak{}consistency} & Zhi Li \\
\texttt{energy/surface-\allowbreak{}energy-\allowbreak{}closure} & Han Wang \\
\texttt{energy/soil-\allowbreak{}heat-\allowbreak{}storage-\allowbreak{}consistency} & Han Wang \\
\texttt{energy/radiation-\allowbreak{}consistency} & Xin Lan \\
\texttt{energy/latent-\allowbreak{}heat-\allowbreak{}et-\allowbreak{}consistency} & Changming Li \\
\texttt{energy/evaporative-\allowbreak{}partition} & Changming Li \\
\texttt{energy/snowmelt-\allowbreak{}energy-\allowbreak{}water} & Siddik Barbhuiya \\
\texttt{momentum/routing-\allowbreak{}conservation} & Zhi Li, Yuanhang Liu \\
\texttt{momentum/routing-\allowbreak{}lag-\allowbreak{}consistency} & Binlan Zhang \\
\texttt{momentum/stage-\allowbreak{}discharge-\allowbreak{}monotonic} & Yuanhang Liu \\
\texttt{momentum/uniform-\allowbreak{}flow-\allowbreak{}friction-\allowbreak{}consistency} & Mofan Zhang \\
\texttt{momentum/froude-\allowbreak{}regime} & Yuanhang Liu \\
\texttt{momentum/wave-\allowbreak{}celerity-\allowbreak{}bounds} & Jingzhi Chen \\
\end{longtable}}

{\footnotesize
\begin{longtable}{@{}P{0.2\linewidth}P{0.72\linewidth}@{}}
\caption{Accepted model proposals and who proposed them. Five accepted proposals count as one probe toward authorship. The model labeled \texttt{lstm} here is listed in the repository under the identifier given in Appendix~\ref{app:repro}.}\label{tab:proposers}\\
\toprule Model & Proposed by \\ \midrule \endfirsthead
\toprule Model & Proposed by \\ \midrule \endhead
\bottomrule \endfoot
\texttt{lstm} & Zhi Li (CU Boulder) \\
\texttt{dhbv2} & Zhi Li (CU Boulder) \\
\texttt{wflow\_\allowbreak{}sbm} & Yuanhang Liu (Independent Researcher) \\
\texttt{summa} & Yuanhang Liu (Independent Researcher) \\
\texttt{cwatm} & Yuanhang Liu (Independent Researcher) \\
\texttt{lisflood} & Yuanhang Liu (Independent Researcher) \\
\texttt{modflow6} & Yaji Wang (University of Illinois Urbana-Champaign) \\
\texttt{hymod} & Xiaoyi Dong (China IPPR International Engineering Co., Ltd., SINOMACH) \\
\texttt{xaj} & Xiaoyi Dong (China IPPR International Engineering Co., Ltd., SINOMACH) \\
\texttt{gr4j} & Yuanhang Liu (Independent Researcher) and Li Zeng (The Hong Kong University of Science and Technology) \\
\texttt{topmodel} & Yan Chen (Colorado State University) \\
\texttt{tank} & Jiao Wang (University of California Davis) \\
\texttt{topkapi} & Jiao Wang (University of California Davis) \\
\texttt{cfe} & Xiaoyi Dong (China IPPR International Engineering Co., Ltd., SINOMACH) \\
\texttt{simhyd} & Haoyang Su (Fudan University and Shanghai Innovation Institute) \\
\texttt{mcp1} & Yuan-Heng Wang (University of Arizona) \\
\texttt{mcp2} & Yuan-Heng Wang (University of Arizona) \\
\texttt{mcp3} & Yuan-Heng Wang (University of Arizona) \\
\texttt{mcp4} & Yuan-Heng Wang (University of Arizona) \\
\texttt{mcp5} & Yuan-Heng Wang (University of Arizona) \\
\texttt{raven} & Yan Chen (Colorado State University) \\
\texttt{hype} & Yan Chen (Colorado State University) \\
\texttt{prms} & Yan Chen (Colorado State University) \\
\texttt{vic} & Valmik Nahata (University of California San Diego) \\
\texttt{mhm} & Valmik Nahata (University of California San Diego) \\
\texttt{shud} & Yuanhang Liu (Independent Researcher) \\
\texttt{noahmp} & Valmik Nahata (University of California San Diego) \\
\texttt{rvic} & Jiao Wang (University of California Davis) \\
\texttt{mizuroute} & Yuanhang Liu (Independent Researcher) \\
\end{longtable}}

\section{Reference models}\label{app:refs}
\Cref{tab:exact} lists the 18 exact references and the number of probes whose gate requires each to pass. \Cref{tab:broken} lists the 59 broken references, the defect built into each, and the probe and criterion that must catch it.

\begin{table}[h]
\centering\small
\caption{Exact reference models and the number of probes whose gate requires them to pass. The first four are the common baselines. The others are written for the probes whose physics they exercise.}\label{tab:exact}
\begin{tabular}{@{}lr@{}}
\toprule Model & Probes requiring a pass \\ \midrule
\texttt{reference\_\allowbreak{}bucket} & 25 \\
\texttt{flex\_\allowbreak{}lumped} & 25 \\
\texttt{sacsma\_\allowbreak{}snow17} & 25 \\
\texttt{flex\_\allowbreak{}topo} & 24 \\
\texttt{reference\_\allowbreak{}coupled} & 3 \\
\texttt{reference\_\allowbreak{}saint\_\allowbreak{}venant} & 2 \\
\texttt{reference\_\allowbreak{}radiative} & 1 \\
\texttt{reference\_\allowbreak{}snow\_\allowbreak{}energy} & 1 \\
\texttt{reference\_\allowbreak{}soil\_\allowbreak{}heat} & 1 \\
\texttt{reference\_\allowbreak{}recharge\_\allowbreak{}order\_\allowbreak{}exact} & 1 \\
\texttt{reference\_\allowbreak{}driven\_\allowbreak{}exchange} & 1 \\
\texttt{reference\_\allowbreak{}evolving\_\allowbreak{}exchange} & 1 \\
\texttt{reference\_\allowbreak{}recharge\_\allowbreak{}exchange} & 1 \\
\texttt{reference\_\allowbreak{}datum\_\allowbreak{}exact} & 1 \\
\texttt{reference\_\allowbreak{}exchange\_\allowbreak{}exact} & 1 \\
\texttt{reference\_\allowbreak{}snyder\_\allowbreak{}router} & 1 \\
\texttt{reference\_\allowbreak{}rating} & 1 \\
\texttt{reference\_\allowbreak{}uniform\_\allowbreak{}flow} & 1 \\
\bottomrule
\end{tabular}
\end{table}
{\footnotesize
\begin{longtable}{@{}P{0.15\linewidth}P{0.36\linewidth}P{0.2\linewidth}P{0.2\linewidth}@{}}
\caption{Deliberately broken reference models (with the \texttt{reference\_} prefix omitted in column 1) and the criterion each is built to trip. Every row is a must-fail check in the acceptance gate. The gate requires each model to fail and to trip its declared criterion. It requires a model to trip no other criterion only in 2 cases, \texttt{snow\_bypass} and \texttt{fixed\_celerity}, where the control exists to separate 1 failure mode from another.}\label{tab:broken}\\
\toprule Reference model & Defect built in & Probe & Criterion \\ \midrule \endfirsthead
\toprule Reference model & Defect built in & Probe & Criterion \\ \midrule \endhead
\bottomrule \endfoot
\texttt{abstraction\_\allowbreak{}blind} & The exact bucket, blind to a prescribed human abstraction. & \texttt{human-\allowbreak{}abstraction} & \texttt{human\_\allowbreak{}abstraction} \\
\texttt{air\_\allowbreak{}emitter} & Negative control for \texttt{energy/radiation-consistency}. & \texttt{radiation-\allowbreak{}consistency} & \texttt{radiative\_\allowbreak{}identity} \\
\texttt{anticipating} & The reference bucket with its runoff smoothed over a centered 7-day window, so every reported value carries 3 days that have not happened yet. & \texttt{causality} & \texttt{causality} \\
\texttt{area\_\allowbreak{}leak} & The bucket losing a share of its runoff that grows with the catchment area it is given: 0\% at 25~km$^2$ and 50\% at 2{,}500~km$^2$. & \texttt{area-\allowbreak{}invariance} & \texttt{invariance} \\
\texttt{calendar} & The bucket model with a recession that drifts with the calendar year. & \texttt{time-\allowbreak{}origin-\allowbreak{}invariance} & \texttt{invariance} \\
\texttt{cheater} & Solves for storage as whatever balances the budget. & \texttt{antecedent-\allowbreak{}monotonicity}, \texttt{catchment-\allowbreak{}closure}, \texttt{precipitation-\allowbreak{}counterfactual}, \texttt{spinup-\allowbreak{}cycle-\allowbreak{}invariance}, \texttt{ungauged-\allowbreak{}basin-\allowbreak{}closure} & \texttt{antecedent\_\allowbreak{}monotonicity}, \texttt{counterfactual\_\allowbreak{}response}, \texttt{state\_\allowbreak{}bounds} \\
\texttt{climatology} & A seasonal climatology baseline: emits the mean runoff and evaporation for the day of the year regardless of the weather, with constant storage. & \texttt{dry-\allowbreak{}down} & \texttt{dry\_\allowbreak{}down} \\
\texttt{constant\_\allowbreak{}lambda} & Fully coupled and correct in every other respect, except that the latent heat of vaporization is a constant instead of a function of temperature. & \texttt{evaporative-\allowbreak{}partition}, \texttt{latent-\allowbreak{}heat-\allowbreak{}et-\allowbreak{}consistency} & \texttt{flux\_\allowbreak{}identity} \\
\texttt{datum\_\allowbreak{}dependent} & Intentionally datum-dependent conductance reference. & \texttt{groundwater-\allowbreak{}datum-\allowbreak{}invariance} & \texttt{datum\_\allowbreak{}flux\_\allowbreak{}invariance} \\
\texttt{degenerate} & The trivial solution: all precipitation evaporates, nothing runs off, storage never changes. & \texttt{catchment-\allowbreak{}closure}, \texttt{precipitation-\allowbreak{}counterfactual}, \texttt{resolution-\allowbreak{}invariance}, \texttt{runoff-\allowbreak{}bounds}, \texttt{spinup-\allowbreak{}cycle-\allowbreak{}invariance}, \texttt{time-\allowbreak{}origin-\allowbreak{}invariance}, \texttt{ungauged-\allowbreak{}basin-\allowbreak{}closure}, \texttt{warming-\allowbreak{}response} & \texttt{counterfactual\_\allowbreak{}response}, \texttt{non\_\allowbreak{}degenerate}, \texttt{response\_\allowbreak{}sign}, \texttt{runoff\_\allowbreak{}bounds} \\
\texttt{degree\_\allowbreak{}day} & The same snowpack, melted on air temperature through a degree-day factor and with a surface energy budget that closes around its evaporation alone. & \texttt{snowmelt-\allowbreak{}energy-\allowbreak{}water} & \texttt{melt\_\allowbreak{}energy} \\
\texttt{diurnal\_\allowbreak{}bias} & The coupled reference's water and energy account with only sensible heat biased: the residual is $+20$~W~m$^{-2}$ from 06:00 to 18:00 local-solar time, and $-20$~W~m$^{-2}$ otherwise. & \texttt{surface-\allowbreak{}energy-\allowbreak{}closure} & \texttt{energy\_\allowbreak{}closure\_\allowbreak{}by\_\allowbreak{}phase} \\
\texttt{energy\_\allowbreak{}leak} & The energy-budget counterpart of \texttt{reference\_leaky}: coherent between water and energy, but 15\% of net radiation never appears in any flux it reports. & \texttt{evaporative-\allowbreak{}partition}, \texttt{latent-\allowbreak{}heat-\allowbreak{}et-\allowbreak{}consistency} & \texttt{energy\_\allowbreak{}closure} \\
\texttt{exchange\_\allowbreak{}sign\_\allowbreak{}error} & A lagged-head bookkeeping reference sharing the smoothed-stage mechanism of \texttt{reference\_exchange\_exact} (declaring \texttt{aquifer\_storage\_coefficient} in \texttt{needs\_static} to match \texttt{probe.yaml}), with a bidirectional river exchange that closes its groundwater balance but reverses a component sign. & \texttt{gw-\allowbreak{}sw-\allowbreak{}exchange-\allowbreak{}consistency} & \texttt{exchange\_\allowbreak{}components} \\
\texttt{fixed\_\allowbreak{}celerity} & Deliberately broken routing reference for \texttt{momentum/wave-celerity-bounds}. & \texttt{wave-\allowbreak{}celerity-\allowbreak{}bounds} & \texttt{wave\_\allowbreak{}celerity\_\allowbreak{}bounds} \\
\texttt{fixed\_\allowbreak{}step} & The reference bucket with its step hard-wired: it claims to run at any resolution and ignores the one it is given, treating every row as a day. & \texttt{resolution-\allowbreak{}invariance} & \texttt{resolution\_\allowbreak{}invariance} \\
\texttt{flat\_\allowbreak{}stage} & The routed bucket that reports a stage at all times equal to a constant. & \texttt{stage-\allowbreak{}discharge-\allowbreak{}monotonic} & \texttt{non\_\allowbreak{}degenerate} \\
\texttt{frozen\_\allowbreak{}soil} & The conductive soil heat reference with reported layer temperature frozen at its initial value. & \texttt{soil-\allowbreak{}heat-\allowbreak{}storage-\allowbreak{}consistency} & \texttt{soil\_\allowbreak{}heat\_\allowbreak{}storage} \\
\texttt{ground\_\allowbreak{}dodge} & Latent heat agrees with the water it evaporated and the energy budget closes, but the sensible heat flux is a fixed share of net radiation that never reads the soil. & \texttt{evaporative-\allowbreak{}partition} & \texttt{partition\_\allowbreak{}shift} \\
\texttt{gw\_\allowbreak{}slow\_\allowbreak{}drift} & An exact bucket whose reporter subtracts 0.001~mm~d$^{-1}$ from runoff and accumulates the deficit in groundwater storage. & \texttt{multi-\allowbreak{}decadal-\allowbreak{}drift} & \texttt{total\_\allowbreak{}storage\_\allowbreak{}drift} \\
\texttt{half\_\allowbreak{}soil} & The conductive soil heat reference with only its reported layer temperature response halved. & \texttt{soil-\allowbreak{}heat-\allowbreak{}storage-\allowbreak{}consistency} & \texttt{soil\_\allowbreak{}heat\_\allowbreak{}storage} \\
\texttt{in\_\allowbreak{}sample} & The bucket model with a leak that only opens outside the range it was built for. & \texttt{extreme-\allowbreak{}event-\allowbreak{}closure} & \texttt{event\_\allowbreak{}water\_\allowbreak{}closure} \\
\texttt{instant\_\allowbreak{}router} & Deliberately wrong rainfall-runoff reference for the routing-lag probe. & \texttt{routing-\allowbreak{}lag-\allowbreak{}consistency} & \texttt{lag\_\allowbreak{}time\_\allowbreak{}bounds} \\
\texttt{inverse\_\allowbreak{}router} & Deliberately wrong geometry-aware reference for the routing-lag probe. & \texttt{routing-\allowbreak{}lag-\allowbreak{}consistency} & \texttt{scaling\_\allowbreak{}monotonicity} \\
\texttt{leaky} & The bucket model with a silent 15\% sink. & \texttt{catchment-\allowbreak{}closure}, \texttt{human-\allowbreak{}abstraction}, \texttt{precipitation-\allowbreak{}counterfactual}, \texttt{spinup-\allowbreak{}cycle-\allowbreak{}invariance}, \texttt{time-\allowbreak{}origin-\allowbreak{}invariance} & \texttt{closure}, \texttt{counterfactual\_\allowbreak{}response} \\
\texttt{leaky\_\allowbreak{}router} & The bucket with a unit hydrograph whose weights sum to 0.999: 0.1\% of every day's runoff enters the channel store and never leaves. & \texttt{routing-\allowbreak{}conservation} & \texttt{routing\_\allowbreak{}conservation} \\
\texttt{no\_\allowbreak{}reflection} & Negative control for \texttt{energy/radiation-consistency}. & \texttt{radiation-\allowbreak{}consistency} & \texttt{radiative\_\allowbreak{}identity} \\
\texttt{noise\_\allowbreak{}sink} & The exact bucket, conserving water internally, which misreports its evaporation by a seeded $\pm 30\%$ and declares the difference as a regional groundwater exchange. & \texttt{exchange-\allowbreak{}response} & \texttt{exchange\_\allowbreak{}response} \\
\texttt{overflowing} & The bucket reporting its runoff plus an unphysical 80\% of the rain. & \texttt{runoff-\allowbreak{}bounds} & \texttt{runoff\_\allowbreak{}bounds} \\
\texttt{overshooting} & The bucket with its runoff sharpened by a derivative term: each day's reported runoff is the bucket's plus 0.8 times the change since the day before, floored at zero. & \texttt{response-\allowbreak{}nonnegativity} & \texttt{response\_\allowbreak{}nonnegativity} \\
\texttt{rating\_\allowbreak{}drift} & The routed bucket with a drifting gauge. & \texttt{stage-\allowbreak{}discharge-\allowbreak{}monotonic} & \texttt{rating\_\allowbreak{}monotonic} \\
\texttt{rating\_\allowbreak{}inverted} & The routed bucket whose rating is read backwards. & \texttt{stage-\allowbreak{}discharge-\allowbreak{}monotonic} & \texttt{rating\_\allowbreak{}loop} \\
\texttt{recharge\_\allowbreak{}blind} & Recharge-ordering reference that ignores \texttt{gw\_recharge}, so its own budget no longer closes. & \texttt{aquifer-\allowbreak{}recharge-\allowbreak{}ordering} & \texttt{aquifer\_\allowbreak{}recharge\_\allowbreak{}ordering} \\
\texttt{recharge\_\allowbreak{}overshoot} & Recharge-ordering reference that sends 120\% of a recharge day to the river, so its storage falls below the control's. & \texttt{aquifer-\allowbreak{}recharge-\allowbreak{}ordering} & \texttt{aquifer\_\allowbreak{}recharge\_\allowbreak{}ordering} \\
\texttt{restless} & The reference bucket with a recession that keeps its own 30-day clock, whatever the weather. & \texttt{spinup-\allowbreak{}cycle-\allowbreak{}invariance}, \texttt{steady-\allowbreak{}state} & \texttt{spinup\_\allowbreak{}cycle\_\allowbreak{}invariance}, \texttt{steady\_\allowbreak{}state} \\
\texttt{saturating} & The reference bucket with its daily runoff capped at 25~mm: exact for ordinary storms, flat for extreme ones. & \texttt{extreme-\allowbreak{}rain} & \texttt{monotone\_\allowbreak{}response} \\
\texttt{shallow\_\allowbreak{}rating} & The exact bucket with a rating borrowed from another reach: its gauge reads Manning normal depth drawn for a section 5 times wider than the one the case declares, so the depth it reports is too shallow for this channel. & \texttt{froude-\allowbreak{}regime} & \texttt{froude\_\allowbreak{}subcritical} \\
\texttt{slow\_\allowbreak{}drift} & An exact bucket whose reporter subtracts 0.012~mm~d$^{-1}$ from runoff and accumulates the deficit in soil storage. & \texttt{multi-\allowbreak{}decadal-\allowbreak{}drift} & \texttt{state\_\allowbreak{}bounds} \\
\texttt{snow\_\allowbreak{}bypass} & Deliberately broken snow model. & \texttt{snowpack-\allowbreak{}mass-\allowbreak{}closure} & \texttt{closure} \\
\texttt{snowless} & Deliberately degenerate snow model. & \texttt{snowpack-\allowbreak{}mass-\allowbreak{}closure} & \texttt{snowpack\_\allowbreak{}response} \\
\texttt{spatial\_\allowbreak{}capacity} & Attribute-dependent fixed internal-capacity fault. & \texttt{ungauged-\allowbreak{}basin-\allowbreak{}closure} & \texttt{state\_\allowbreak{}bounds} \\
\texttt{spatial\_\allowbreak{}et} & Causal attribute-dependent ET fault. & \texttt{ungauged-\allowbreak{}basin-\allowbreak{}closure} & \texttt{et\_\allowbreak{}plausible} \\
\texttt{spatial\_\allowbreak{}forcing} & Causal attribute-dependent forcing fault. & \texttt{ungauged-\allowbreak{}basin-\allowbreak{}closure} & \texttt{forcing\_\allowbreak{}fidelity} \\
\texttt{spatial\_\allowbreak{}frozen} & Causal attribute-dependent frozen fault. & \texttt{ungauged-\allowbreak{}basin-\allowbreak{}closure} & \texttt{non\_\allowbreak{}degenerate} \\
\texttt{spatial\_\allowbreak{}gain} & Causal attribute-dependent gain fault. & \texttt{ungauged-\allowbreak{}basin-\allowbreak{}closure} & \texttt{closure} \\
\texttt{spatial\_\allowbreak{}loss} & Causal attribute-dependent loss fault. & \texttt{ungauged-\allowbreak{}basin-\allowbreak{}closure} & \texttt{closure} \\
\texttt{spatial\_\allowbreak{}negative} & Causal attribute-dependent negative fault. & \texttt{ungauged-\allowbreak{}basin-\allowbreak{}closure} & \texttt{non\_\allowbreak{}degenerate} \\
\texttt{steady\_\allowbreak{}sink} & \texttt{reference\_token\_exchange} with its noise removed: evaporation overstated by a constant 0.3~mm~d$^{-1}$ (13\% of precipitation over the record), declared as a steady regional inflow so the budget closes, with a head response $10^6$ times weaker than the positive controls. & \texttt{exchange-\allowbreak{}response} & \texttt{exchange\_\allowbreak{}response} \\
\texttt{streamflow\_\allowbreak{}only} & Predicts discharge and nothing else, like most published rainfall-runoff models. & \texttt{warming-\allowbreak{}response} & \texttt{response\_\allowbreak{}sign} \\
\texttt{stuck\_\allowbreak{}router} & The bucket with a unit hydrograph whose weights sum to 0.9: 10\% of every day's runoff enters the channel store and never leaves. & \texttt{routing-\allowbreak{}conservation} & \texttt{routing\_\allowbreak{}conservation} \\
\texttt{sublimating} & The bucket losing 40\% of every snowfall to unreported sublimation. & \texttt{phase-\allowbreak{}counterfactual} & \texttt{phase\_\allowbreak{}invariance} \\
\texttt{sublimation\_\allowbreak{}blind} & Fully coupled with a temperature-dependent latent heat, but it converts snow sublimation at the latent heat of vaporization instead of sublimation. & \texttt{latent-\allowbreak{}heat-\allowbreak{}et-\allowbreak{}consistency} & \texttt{flux\_\allowbreak{}identity} \\
\texttt{thirsty} & The bucket with evaporation that ignores demand: a fixed 2\% of the soil store leaves every day, capped only by available water. & \texttt{pet-\allowbreak{}consistency} & \texttt{demand\_\allowbreak{}consistency} \\
\texttt{token\_\allowbreak{}exchange} & \texttt{reference\_noise\_sink} with 1 line added: a head-driven trickle $10^6$ times weaker than the positive controls, folded into the declared exchange and soil storage so the budget closes. & \texttt{exchange-\allowbreak{}response} & \texttt{exchange\_\allowbreak{}response} \\
\texttt{two\_\allowbreak{}head} & A model with decoupled water and energy heads that close their individual budgets without agreeing on latent heat. & \texttt{evaporative-\allowbreak{}partition}, \texttt{latent-\allowbreak{}heat-\allowbreak{}et-\allowbreak{}consistency} & \texttt{flux\_\allowbreak{}identity}, \texttt{partition\_\allowbreak{}shift} \\
\texttt{warming\_\allowbreak{}free} & The same snowpack, melting on the energy actually available and reporting its cold content honestly, but charging its surface budget for the fusion alone. & \texttt{snowmelt-\allowbreak{}energy-\allowbreak{}water} & \texttt{melt\_\allowbreak{}energy} \\
\texttt{wrong\_\allowbreak{}roughness} & Deliberately inconsistent steady reach. & \texttt{uniform-\allowbreak{}flow-\allowbreak{}friction-\allowbreak{}consistency} & \texttt{uniform\_\allowbreak{}flow\_\allowbreak{}friction} \\
\texttt{wrong\_\allowbreak{}slope} & Deliberately inconsistent steady reach. & \texttt{uniform-\allowbreak{}flow-\allowbreak{}friction-\allowbreak{}consistency} & \texttt{uniform\_\allowbreak{}flow\_\allowbreak{}friction} \\
\texttt{zero\_\allowbreak{}exchange} & Intentionally frozen zero-exchange reference. & \texttt{groundwater-\allowbreak{}datum-\allowbreak{}invariance} & \texttt{datum\_\allowbreak{}flux\_\allowbreak{}invariance} \\
\end{longtable}}

\section{Tolerance and seed sensitivity}\label{app:tolerance}
\Cref{tab:sweep} varies the closure tolerance over 4 orders of magnitude and counts the evaluated models that would fail. The residual of the worst model, \code{xaj}, is 2.5\%, and the verdict on the broken reference is unchanged for every tolerance below 15\%. \Cref{tab:seeds} shows the seed sensitivity that motivates running several seeds. The 20 seeds for \code{catchment-closure} and the 10 seeds for \code{extreme-rain} were drawn at random from a recorded generator, and the seed lists are included with the source of this article.

\begin{table}[t]
\centering\small
\caption{Sensitivity of the closure verdicts to the tolerance. Entries are the number of the 22 evaluated models whose whole-record residual on \code{mass/catchment-closure} exceeds each limit. The broken reference \code{reference\_leaky} has a residual of 15\% by construction.}\label{tab:sweep}
\begin{tabular}{@{}rcc@{}}
\toprule
Tolerance (\% of precipitation) & Evaluated models failing (of 22) & \code{reference\_leaky} \\ \midrule
0.01 & 4 & fails \\
0.1 & 2 & fails \\
0.5 & 1 & fails \\
1 & 1 & fails \\
2 & 1 & fails \\
5 (adopted) & 0 & fails \\
10 & 0 & fails \\
20 & 0 & passes \\
\bottomrule
\end{tabular}
\end{table}

\begin{table}[t]
\centering\small
\caption{Seed sensitivity of 2 probes. Top: \code{mass/catchment-closure} on 20 random seeds. Bottom: \code{mass/extreme-rain} on 10 random seeds. The column on the right gives the criterion trips over the seeds, or the share of added rain that runs off over the ladder up to 10 times the largest storm, as the range over seeds.}\label{tab:seeds}
\begin{tabular}{@{}P{0.14\linewidth}P{0.21\linewidth}cP{0.38\linewidth}@{}}
\toprule
Probe & Model & Seeds passed & Criterion trips or measured range \\ \midrule
\code{catchment-closure} & \code{reference\_bucket} & 20/20 & largest residual $2.6\times10^{-16}$ of precipitation \\
\code{catchment-closure} & \code{flex\_lumped} & 20/20 & largest residual $2.2\times10^{-15}$ of precipitation \\
\code{catchment-closure} & \code{sacsma\_snow17} & 20/20 & largest residual $2.0\times10^{-15}$ of precipitation \\
\code{catchment-closure} & \code{reference\_leaky} & 0/20 & \code{closure} 20 of 20 \\
\code{catchment-closure} & \code{reference\_cheater} & 0/20 & \code{state\_bounds} 20 of 20 \\
\code{catchment-closure} & \code{reference\_degenerate} & 0/20 & \code{non\_degenerate} 20, \code{et\_plausible} 18 of 20 \\
\midrule
\code{extreme-rain} & \code{reference\_bucket} & 10/10 & share 0.56 to 1.00 \\
\code{extreme-rain} & \code{reference\_saturating} & 4/10 & share 0.01 to 0.22 \\
\bottomrule
\end{tabular}
\end{table}

\section{Evaluated models and reproducibility}\label{app:models}

\subsection{Adapting a model}
The repository documents 6 steps for adding a model. First, the proposer opens a model proposal and the maintainer accepts it. Second, the proposer decides which variables the model honestly emits. Third, an adapter translates between the contract and the native model. Fourth, a Dockerfile or a native entrypoint makes the run reproducible. Fifth, a contract check, \code{ht verify-adapter}, runs 1 seed on the evaluation window before any physics is scored. Sixth, the contributor runs the suite with \code{ht run} and submits the manifest, the adapter and the archived results. The common failure modes are padding the list of emitted variables, which turns an honest limit into a false claim, and declaring a forcing input that the adapter does not consume.

\subsection{Standing of the evaluated models}
\Cref{tab:standings} lists the probes passed over probes scored for every model, split by law.

{\footnotesize
\begin{longtable}{@{}llcccc@{}}
\caption{Standing of the 32 evaluated models in suite 0.1.0. Entries are probes passed over probes scored. A probe is not scored when the model does not report the variables it needs or cannot run at the probe's time step (N/A). Rows are ordered by class and then by the number of probes passed. The LSTM is the model behind the case study in Section~\ref{sec:case}.}\label{tab:standings}\\
\toprule Model & Class & All & Mass & Energy & Momentum \\ \midrule \endfirsthead
\toprule Model & Class & All & Mass & Energy & Momentum \\ \midrule \endhead
\bottomrule \endfoot
\texttt{flex\_\allowbreak{}lumped} & Physical reference & 25/25 & 19/19 & 1/1 & 5/5 \\
\texttt{sacsma\_\allowbreak{}snow17} & Physical reference & 25/25 & 20/20 & 1/1 & 4/4 \\
\texttt{flex\_\allowbreak{}topo} & Physical reference & 24/24 & 19/19 & 1/1 & 4/4 \\
\texttt{mcp1} & AI-based & 18/20 & 17/19 & 1/1 & -- \\
\texttt{mcp3} & AI-based & 17/21 & 16/19 & 1/1 & 0/1 \\
\texttt{mcp2} & AI-based & 15/20 & 15/19 & 0/1 & -- \\
\texttt{dhbv2} & AI-based & 15/21 & 13/19 & 1/1 & 1/1 \\
\texttt{mcp4} & AI-based & 14/20 & 13/19 & 1/1 & -- \\
\texttt{mcp5} & AI-based & 14/20 & 14/19 & 0/1 & -- \\
\texttt{lstm} & AI-based & 5/11 & 5/11 & -- & -- \\
\texttt{mhm} & Process-based & 21/22 & 19/20 & 1/1 & 1/1 \\
\texttt{gr4j} & Process-based & 20/21 & 18/19 & 1/1 & 1/1 \\
\texttt{hymod} & Process-based & 20/21 & 18/19 & 1/1 & 1/1 \\
\texttt{topkapi} & Process-based & 20/21 & 18/19 & 1/1 & 1/1 \\
\texttt{prms} & Process-based & 20/22 & 18/20 & 1/1 & 1/1 \\
\texttt{tank} & Process-based & 19/20 & 18/19 & 1/1 & -- \\
\texttt{lisflood} & Process-based & 19/21 & 17/19 & 1/1 & 1/1 \\
\texttt{xaj} & Process-based & 19/21 & 17/19 & 1/1 & 1/1 \\
\texttt{topmodel} & Process-based & 19/22 & 17/19 & 1/1 & 1/2 \\
\texttt{wflow\_\allowbreak{}sbm} & Process-based & 19/22 & 16/19 & 1/1 & 2/2 \\
\texttt{cwatm} & Process-based & 17/21 & 16/19 & 0/1 & 1/1 \\
\texttt{noahmp} & Process-based & 17/25 & 14/20 & 3/5 & -- \\
\texttt{vic} & Process-based & 16/23 & 15/19 & 1/4 & -- \\
\texttt{hype} & Process-based & 15/23 & 13/20 & 1/1 & 1/2 \\
\texttt{simhyd} & Process-based & 11/20 & 11/19 & 0/1 & -- \\
\texttt{shud} & Process-based & 9/22 & 8/19 & 0/1 & 1/2 \\
\texttt{summa} & Process-based & 9/23 & 7/19 & 1/3 & 1/1 \\
\texttt{raven} & Process-based & 7/22 & 6/20 & 0/1 & 1/1 \\
\texttt{modflow6} & Process-based & 3/3 & 3/3 & -- & -- \\
\texttt{cfe} & Process-based & 1/1 & 1/1 & -- & -- \\
\texttt{mizuroute} & Process-based & 1/1 & -- & -- & 1/1 \\
\texttt{rvic} & Process-based & 0/1 & -- & -- & 0/1 \\
\end{longtable}}

\subsection{Reproducibility}\label{app:repro}
All numbers in this article derive from suite 0.1.0, archived on Zenodo, and from the results archive in the repository. The acceptance gate of every probe was re-run on 2 November 2026 with the command \code{ht gate}, and the standings of \cref{tab:standings} reproduce from the archive. The models are listed in the archive under their repository identifiers. The LSTM of this article appears there as \code{google\_flood\_forecast}, version 0.1.0-828dfc5.2, with released weights, and it was run in a Docker container on 8 CPU cores. The mass-conserving perceptron checkpoints \code{mcp1} to \code{mcp5} correspond to epochs 27, 3, 619, 2707 and 2999 of 1 architecture. Versions of all other models are recorded in the archive.

\end{document}